\documentclass[a4paper,11pt]{article}
\pdfoutput=1
\usepackage{jheppub}
\usepackage[utf8]{inputenc}

\usepackage{lmodern}

\usepackage{amsfonts}
\usepackage{amssymb}
\usepackage{mathrsfs}
\usepackage{mathtools}
\usepackage{mleftright}

\DeclareFontFamily{OT1}{pzc}{}
\DeclareFontShape{OT1}{pzc}{m}{it}{<-> s * [1.200] pzcmi7t}{}
\DeclareMathAlphabet{\mathpzc}{OT1}{pzc}{m}{it}

\newcommand*{\xhat}[1]{#1\kern-0.5em\hat{\phantom{#1}}}

\renewcommand{\Re}{\operatorname{Re}}
\renewcommand{\Im}{\operatorname{Im}}
\DeclareMathOperator{\tr}{tr}
\DeclareMathOperator{\vol}{vol}
\DeclareMathOperator{\rk}{rk}
\DeclareMathOperator*{\Res}{Res}

\def\A{a}
\def\TT{\mathcal{O}_{T\bar{T}}}
\def\X{X}
\def\Y{Y}
\def\W{W}
\def\mflux{\boldsymbol{\mathfrak{m}}}
\def\V{\mathbf{V}}
\newcommand{\smod}[1]{\langle #1 \rangle}
\newcommand{\EpsZ}[2]{\mathrm{Z}\left|\begin{smallmatrix}#1\\#2\end{smallmatrix}\right|}
\newcommand{\EpsZreg}[2]{\mathrm{Z}^{\text{reg}}\left|\begin{smallmatrix}#1\\#2\end{smallmatrix}\right|}

\def\fint{{=}\kern-1em\int}

\usepackage{tikz}

\usetikzlibrary{
    decorations.markings,
    decorations.pathmorphing
}

\def\graphwidth{2.0}
\def\graphheight{0.8}
\def\cutval{0.7}
\def\markheight{0.05}
\def\contourgap{0.12}
\def\axescolor{black}
\def\contourcolor{red}

\tikzset{cutstyle/.style={decorate, decoration={zigzag, segment length=6, amplitude=2}, draw=black}}
\tikzset{arrow data/.style 2 args={decoration={markings, mark=at position #1 with \arrow{#2}}, postaction=decorate}}

\newcommand\drawdot[3]{
    \draw[fill] #1 circle (.7pt)
    node[black, shift=#2] {#3};
}
\newcommand\drawbigdot[3]{
    \draw[blue, fill=blue] #1 circle (1pt)
    node[blue, shift=#2] {#3};
}

\newcommand\drawxmark[3]{
    \draw[-]
    #1+(+\markheight, +\markheight) -- #1
    #1+(+\markheight, -\markheight) -- #1
    #1+(-\markheight, +\markheight) -- #1
    #1+(-\markheight, -\markheight) -- #1
    #1 node[black, shift=#2] {#3};
}

\newcommand\lateralborelym{
    \begin{tikzpicture}[scale=2, thick]
    \draw[-latex, \axescolor] (-\graphwidth, 0) -- (+\graphwidth, 0) node[right] {$\Re\zeta$};
    \draw[-latex, \axescolor] (0, -\graphheight) -- (0, +\graphheight) node[below left] {$\Im\zeta$};
    \drawdot{(-\cutval,0)}{(-60:0.5)}{$-1$};
    \draw[cutstyle] (-\cutval, 0) -- ({-\graphwidth+0.1}, 0);
    \draw[\contourcolor, ultra thick, arrow data={0.3}{latex}, arrow data={0.645}{latex}, arrow data={0.95}{latex}]
        (0,0) -- (-\graphwidth,-{1.5*\contourgap}) node[shift=(-60:0.6)] {$\mathcal{S}_{-\pi}$}
        (0,0) -- (-\graphwidth,+{1.5*\contourgap}) node[shift=(+60:0.6)] {$\mathcal{S}_{+\pi}$}
        (0,0) -- ({+\graphwidth-0.2},0) node[shift=(+120:0.6)] {$\mathcal{S}_{0}$};
    \end{tikzpicture}
}

\newcommand\contourfigure{
    \begin{tikzpicture}[scale=2, thick]
    \draw[-latex, \axescolor] (-2*\graphwidth/3, 0) -- (+2*\graphwidth/3, 0) node[right] {$\Re u$};
    \draw[-latex, \axescolor] (0, -\graphheight) -- (0, +\graphheight);
    \drawdot{(0,0)}{(-120:0.5)}{$0$};
    \drawxmark{(-\graphwidth/6,0)}{(90:0.5)}{$-y$};
    \draw (\graphwidth/3,0) node[red, shift=(-60:0.5)] {$\varrho$};
    \draw[cutstyle] (0, 0) -- (+15*\graphwidth/24, 0);
    \draw[\contourcolor, ultra thick, arrow data={0.15}{latex}, arrow data={0.25}{latex}, arrow data={0.58}{latex}, arrow data={0.75}{latex}]
        (0,\contourgap) arc (90:270:\contourgap)
        (\graphwidth/3,-\contourgap) -- (0,-\contourgap)
        (0,\contourgap) -- (\graphwidth/3,\contourgap)
        ([shift={(10:\graphwidth/3)}]0,0) arc (10:350:\graphwidth/3);
    \draw (\graphwidth/8,0) node[red, shift=(90:0.6)] {${\gamma}_{\mathrm{cut}}$};
    \draw (-\graphwidth/3,-\graphwidth/3) node[red, shift=(45:0.2)] {${\gamma}_{\mathrm{circ}}$};
    \filldraw [fill=white, draw=white] (-0.45,0.55) rectangle (-0.05,0.75);
    \draw (0, +\graphheight) node[below left] {$\Im u$};
    \end{tikzpicture}
}

\newcommand\saddlenegfigure{
    \begin{tikzpicture}[scale=2, thick]
    \coordinate (saddle) at (-\graphwidth/5,0);
    \coordinate (vs) at (-4.9*\graphwidth/6,0);
    \draw[-latex, \axescolor] (-\graphwidth, 0) -- (+2*\graphwidth/3, 0) node[right] {$\Re v$};
    \draw[-latex, \axescolor] (0, -\graphheight) -- (0, +\graphheight) node[below right] {$\Im v$};
    \drawdot{(0,0)}{(-135:0.4)}{$0$};
    \draw[cutstyle] (0, 0) -- (+15*\graphwidth/24, 0);
    \draw[\contourcolor, ultra thick, arrow data={0.4}{latex}, arrow data={0.9}{latex}]
        (saddle) arc (0:360:5*\graphheight/6);
    \drawxmark{(vs)}{(-30:0.7)}{$v_{\mathrm{sing}}$};
    \drawbigdot{(saddle)}{(150:0.7)}{$-\omega_\star^2$};
    \draw (-\graphwidth/3,-\graphwidth/3) node[red, shift=(0:0.3)] {$\gamma_<$};
    \end{tikzpicture}
}

\newcommand\saddleposfigure{
    \begin{tikzpicture}[scale=2, thick]
    \coordinate (saddle) at (0,-2*\graphheight/3);
    \coordinate (wp) at (+\graphwidth/2,0);
    \coordinate (wn) at (-\graphwidth/2,0);
    \draw[-latex, \axescolor] (-2*\graphwidth/3, 0) -- (+2*\graphwidth/3, 0) node[right] {$\Re w$};
    \draw[-latex, \axescolor] (0, -\graphheight) -- (0, +\graphheight) node[below right] {$\Im w$};
    \draw[\contourcolor, ultra thick, arrow data={0.4}{latex}, arrow data={0.7}{latex}]
                (wn) to[out=0,in=180] (saddle)
                (saddle) to[out=0,in=180] (wp);
    \draw[\contourcolor, dashed, ultra thick, arrow data={0.4}{latex}, arrow data={0.7}{latex}]
        (wn) -- (wp);
    \drawxmark{(wn)}{(90:0.5)}{$-w_{\mathrm{sing}}$};
    \drawxmark{(wp)}{(90:0.5)}{$+w_{\mathrm{sing}}$};
    \drawbigdot{(saddle)}{(-20:0.7)}{$-\mathrm{i}\omega_\star$};
    \draw (-\graphwidth/3,-\graphwidth/4) node[red, shift=(0:0.3)] {$\gamma_>$};
    \end{tikzpicture}
}

\title{\huge\boldmath Exact $T\bar{T}$-deformation of two-dimensional Yang--Mills theory on the sphere}

\author[a]{Luca Griguolo,}
\author[b]{Rodolfo Panerai,}
\author[a]{Jacopo Papalini,}
\author[c]{and Domenico Seminara}

\affiliation[a]{Dipartimento SMFI, Universit\`a di Parma and INFN Gruppo Collegato di Parma, Viale G.P. Usberti 7/A, 43100 Parma, Italy}
\affiliation[b]{Department of Physics and Astronomy, Uppsala University, Box 516, SE-75120 Uppsala, Sweden}
\affiliation[c]{Dipartimento di Fisica, Universit\`a di Firenze and INFN Sezione di Firenze, via G. Sansone 1, 50019 Sesto Fiorentino, Italy} 

\emailAdd{luca.griguolo@unipr.it}
\emailAdd{rodolfo.panerai@physics.uu.se}
\emailAdd{jacopo.papalini@unipr.it}
\emailAdd{seminara@fi.infn.it}

\abstract{
We study the $T\bar{T}$ deformation of two-dimensional Yang--Mills theory at genus zero by carrying out the analysis at the level of its instanton representation.
We first focus on the perturbative sector by considering its power expansion in the deformation parameter $\mu$. By studying the resulting asymptotic series through resurgence theory, we determine the nonperturbative contributions that enter the result for $\mu<0$.
We then extend this analysis to any flux sector by solving the relevant flow equation. Specifically, we impose boundary conditions corresponding to two distinct regimes: the quantum undeformed theory and the semiclassical limit of the deformed theory.
The full partition function is obtained as a sum over all magnetic fluxes.
For any $\mu>0$, only a finite portion of the quantum spectrum survives and the partition function reduces to a sum over a finite set of representations.
For $\mu<0$, nonperturbative contributions regularize the partition function through an intriguing mechanism that generates nontrivial subtractions.
}

\begin{document} 
\maketitle
\flushbottom

\section{Introduction}
The $T\bar{T}$ deformation \cite{Smirnov:2016lqw,Cavaglia:2016oda} is a deformation of local relativistic quantum field theories in two dimensions induced by a specific irrelevant local operator, quadratic in the stress-energy tensor. This operator is unambiguously defined in the presence of translational invariance since its point-splitted version has a regular pinching limit, up to total derivatives \cite{Zamolodchikov:2004ce}.
The deformation generates a one-parameter family of quantum field theories with strongly-coupled dynamics at high energies, and despite being in general expected to destroy short-distance locality, it exhibits remarkable properties. It preserves many of the symmetries of the original theory, and it is amenable to exact computations. For instance, the finite-volume spectrum of the deformed theory is described by a differential equation of Burgers type \cite{Smirnov:2016lqw, Cavaglia:2016oda}. The surprising amount of control that the deformation allows seems to provide a consistent way to move against the renormalization-group flow and explore unconventional fixed points in the ultraviolet.

Being triggered by the stress-energy tensor, the deformation appears to be rooted in geometry. In fact, it can equivalently be formulated as a coupling to topological gravity \cite{Dubovsky:2017cnj,Dubovsky:2018bmo,Tolley:2019nmm} or random background metrics \cite{Cardy:2018sdv}.
Much of the literature on the $T\bar{T}$ deformation deals with its application to conformal field theories, where the geometric dependence of the undeformed spectrum is fixed by conformal invariance. In this context, the action of the deformation has been observed leading to radically different regimes according to the sign of the irrelevant coupling $\mu$.
For a positive sign, the density of states of a deformed conformal field theory interpolates between the typical Cardy growth and a Hagedorn-like growth \cite{Giveon:2017nie} signaling nonlocal features of the deformed field theory in the UV, reminiscent of a stringy behavior.
For a negative sign, the spectrum seemingly undergoes a partial complexification \cite{Aharony:2018bad}, putting into question the consistency of the theory at finite volume. The presence of nonperturbative effects in the deformation parameter has been advocated \cite{Aharony:2018bad} to cure this pathological behavior, although we are not aware of any precise computation in this direction.
$T\bar{T}$-deformed conformal field theories with negative $\mu$ have also been suggested leading to an extension of the holographic dictionary \cite{McGough:2016lol,Giveon:2017nie,Kraus:2018xrn,Chakraborty:2019mdf}, potentially describing quantum gravity confined in a portion of the AdS$_3$ bulk of radius $r_{\mathrm{c}} \propto 1/\sqrt{-\mu}$.

The present paper deals with the deformation of gauge theories.
Pure Yang--Mills theory in two dimensions is quite different from its higher-dimensional counterparts in that it does not allow for propagating degrees of freedom. The theory is invariant under a large group of spacetime symmetries that make the dependence on the geometry almost trivial and render the theory solvable \cite{Rusakov:1990rs,Witten:1991we} (see \cite{Cordes:1994fc} for a review on the subject). In the context of the $T\bar{T}$ deformation, Yang--Mills theory was studied in \cite{Conti:2018jho,Ireland:2019vvj,Santilli:2020qvd,Pavshinkin:2021jpy} and its large-$N$ limit was explored in \cite{Santilli:2018xux,Gorsky:2020qge}.
In the present work, we continue the study of $T\bar{T}$-deformed two-dimensional Yang--Mills theory that we initiated in \cite{Griguolo:2022xcj} with the analysis of the abelian case. Specifically, we focus on the $\mathrm{U}(N)$ gauge theory at genus zero, although much of our results can be generalized to arbitrary groups and topologies.

There are two main features of the deformed theory that we set out to address. These are associated with the two different sign choices for the deformation parameter $\mu$. For $\mu>0$, only a finite number of states in the deformed spectrum can be accessed by solving the relevant flow equation, the rest of the spectrum lying behind a divergence. If one insists on preserving the hierarchy of states of the undeformed theory, one should postulate that an infinite number of energy levels should decouple from the theory. For $\mu<0$, to obtain a well-defined partition function, one should incorporate instanton-like corrections in $\mu$ whose precise form is determined by imposing appropriate physical requirements.

\paragraph{Summary of results.}
In order to find a dynamical explanation for both features, we construct the deformed partition function for each flux sector $\mathpzc{z}_{\mflux}$. These are sectors of the theory associated with stationary points of the classical action, labeled by the quantized magnetic flux vector $\mflux\in\mathbb{Z}^N$. To determine the correct solutions of the differential equation describing the $T\bar{T}$ flow, 
\begin{align}\label{EQ:flow_equation_intro}
    \frac{\partial \mathpzc{z}_{\mflux}}{\partial \mu} + 2a \, \frac{\partial^2\!\mathpzc{z}_{\mflux}}{\partial a^2} = 0 \;,
\end{align}
($a$ denotes the total area) we must impose suitable boundary conditions. These correspond to the two important physical regimes we have access to. The first is the undeformed theory in its fully-quantum regime. The second is the deformed theory in its semiclassical limit.

This approach is essential for two reasons. For $\mu>0$, the truncation of the spectrum is associated with nonanalyticities of the partition function. In the abelian theory \cite{Griguolo:2022xcj}, these were interpreted in terms of an infinite number of infinite-order quantum phase transitions. This peculiar behavior only emerges when taking the sum over $\mflux$: each $\mathpzc{z}_{\mflux}$ is, in fact, analytic when $\mu>0$. On the other hand, for $\mu<0$, we need to fix nonperturbative terms to which the undeformed limit $\mu\to0$ is insensitive. Crucially, these are relevant for the semiclassical regime of the theory that can be probed by taking a specific double scaling limit. Only by having the result for the deformed $\mathpzc{z}_{\mflux}$ can we match its semiclassical limit against the deformed action evaluated on the associated classical saddle.

In analogy with the undeformed case, the partition function can be expressed through a sum over inequivalent irreducible representations of the gauge group. Schematically, we find that the full deformed partition function, written in terms of the deformed Hamiltonian
\begin{align}
    H = \frac{g_{\text{YM}}^2\,C_2(R)/2}{1-\mu g_{\text{YM}}^2\,C_2(R)} \;,
\end{align}
reads
\begin{align}
    Z
    &= \sum_{R|H>0} (\dim R)^{2} \; e^{-aH} & &\text{for $\mu>0$,}
    \label{EQ:Z_deformed_positive_intro} \\
    Z
    &= \sum_{\mathmakebox[\widthof{$\scriptstyle{R|H>0}$}][c]{R}} (\dim R)^{2} \, \bigg(e^{-aH} - \sum_{k=0}^{k_{\text{max}}}\lambda_k\bigg) + \mathcal{R} & &\text{for $\mu<0$.}
    \label{EQ:Z_deformed_negative_intro}
\end{align}
For $\mu>0$, the sum extends over the finite number of representations $R$ for which $H>0$. For $\mu<0$, the sum is unrestricted, though in order for it to converge a finite number of terms $\lambda_k$ is subtracted. These are the first few terms in the $\mu$-expansion of $e^{-aH}$, and the upper bound $k_{\text{max}}$ is the minimum value for which the sum over $R$ converges. Each of the $\lambda_k$ carries a factor of $e^{a/(2\mu)}$, making each term nonperturbative in $\mu$. The same factor appears in the residual term $\mathcal{R}$, which is itself a solution of the flow equation \eqref{EQ:flow_equation_intro}.

\paragraph{Outlook.}
There is ample reason to believe that \eqref{EQ:Z_deformed_positive_intro} should directly generalize to arbitrary gauge groups and manifolds, with and without boundaries, in analogy with the case of $\mu=0$. This is necessary in order to preserve the topological composition properties of the undeformed theory. In fact, these rely solely on the orthogonality of characters and are unaffected by the deformation of the Hamiltonian or by the finite range of the sum.

The situation is quite different for the partition function \eqref{EQ:Z_deformed_negative_intro}. Specifically, it is less obvious how $\mathcal{R}$ should be modified to account for different groups and topologies. Furthermore, the nonperturbative terms appear to be incompatible with the gluing rules of undeformed Yang--Mills theory, at least in their simplest form. This inconsistency could be interpreted by invoking a breakdown of locality in the $\mu<0$ regime. It would be interesting to investigate these points further, e.g.\ on the torus\footnote{
    For the undeformed torus partition function in a generic flux sector, see \cite{Griguolo:1998kq,Griguolo:2001ce,Griguolo:2004jp}
}

Furthermore, it is natural to employ our analysis of the flow equation to extend previous results at large $N$ \cite{Santilli:2018xux,Gorsky:2020qge}. In particular, it should be possible to obtain the full $1/N$ expansion of the deformed theory by studying the differential equation governing the deformation of the free energy in the large-$N$ limit. We hope to report on this point in the future.

\paragraph{Outline of the paper.} The paper is organized as follows. In Section~\ref{SEC:YM_theory}, we briefly review relevant aspects of undeformed Yang--Mills theory in two dimensions. In Section~\ref{SEC:TT_deformed_YM_theory}, we introduce its $T\bar{T}$ deformation both at the level of the deformed Lagrangian and in terms of a flow equation for the partition function. We then discuss the subtleties that arise for both sign choices of the deformation parameter. In Section~\ref{SEC:z0}, we construct the deformed zero-flux sector by Borel resumming the associated power expansion in the deformation parameter. The analytic properties of the associated Borel transform signal the presence of nonperturbative contributions at $\mu<0$. We determine the form of such terms with resurgence theory. In Section~\ref{SEC:zm}, we compute the partition function for arbitrary flux sectors by solving the relevant flow equation. To reproduce the correct undeformed limit, we project the initial condition on a complete set of solutions using the Ramanujan master theorem. In Section~\ref{SEC:Z_full}, we sum over all flux sectors to obtain the explicit form of the full deformed partition function. It involves the use of the multidimensional Poisson summation formula and certain generalizations thereof. In Section~\ref{SEC:semiclassical_limit}, we show how the deformed flux sectors obtained in Section~\ref{SEC:zm} reproduce the correct semiclassical limit, thus confirming our choice of nonperturbative corrections. We argue that the truncation of the spectrum is due to destructive interference between deformed flux sectors. Two technical appendices complete the manuscript.

\section{Yang--Mills theory in two dimensions}\label{SEC:YM_theory}
\subsection{General properties}
Let us consider a Euclidean gauge theory on a compact orientable Riemann surface $\Sigma$ of genus $\mathbf{g}$. We denote the gauge group and its Lie algebra with $G$ and $\mathfrak{g}$, respectively. In our conventions, the gauge fields are hermitian, and we define the curvature in terms of the gauge connection $A$ as $F = \mathrm{d}A - \mathrm{i}A \wedge A$.

The action of pure Yang--Mills theory
\begin{align}\label{EQ:S_YM_with_F}
    S_{\text{YM}} = \frac{1}{2g^2_{\text{YM}}}\int_\Sigma \tr F\wedge\star F
\end{align}
can be rewritten in terms of a single $\mathfrak{g}$-valued scalar $f = \star F$ as
\begin{align}\label{EQ:S_YM_with_f}
    S_{\text{YM}} = \frac{1}{2g^2_{\text{YM}}}\int_\Sigma \eta \, \tr f^2 \;,
\end{align}
where $\eta$ is the volume form on $\Sigma$. An alternative action for the theory can be obtained by introducing an auxiliary $\mathfrak{g}$-valued scalar $\phi$,
\begin{align}\label{EQ:S_top}
    S_{\text{top}} = \mathrm{i}\int_\Sigma \tr(\phi F) + \frac{g^2_{\text{YM}}}{2} \int_\Sigma \eta \, \tr\phi^2 \;.
\end{align}
This last expression shows that the theory is invariant under a large group of local symmetries, known as \emph{area-preserving diffeomorphisms} \cite{Cordes:1994fc}. As a consequence, the partition function is sensitive to the underlying geometry only through the total area $\A = \vol\Sigma$. In fact, since the action is invariant under an appropriate simultaneous rescaling of $\A$ and of the Yang--Mills coupling $g_{\text{YM}}$, the dependence on such couplings comes only through the combination $g^2_{\text{YM}}\A$.

The full quantum theory is solvable \cite{Rusakov:1990rs,Witten:1991we}.
Its partition function can be written as a sum over inequivalent irreducible representations of the gauge group,
\begin{align}\label{EQ:Z_YM}
    Z
    &= \sum_{R} (\dim R)^{2-2\mathbf{g}} \, e^{-g^2_{\text{YM}}\A\,C_2(R)/2} \;,
\end{align}
where $C_2(R)$ is the eigenvalue of the quadratic Casimir of the representation $R$.

A localization argument \cite{Witten:1992xu} leads to an alternative representation for the partition function as a sum over solutions of the Yang--Mills equation (i.e.\ solutions of $\mathrm{D} f = 0$).
For $\mathbf{g} = 0$, these unstable instantons are labeled by their associated GNO-quantized magnetic flux
\begin{align}
    \mflux = \frac{1}{2\pi} \int_\Sigma F
\end{align}
which belongs to $\Lambda_G$, the cocharacter lattice of $G$. In other words, $\mflux$ is an element of the Lie algebra of the maximal torus $H$ of $G$ such that $e^{2\pi\mathrm{i}\mflux} = \mathbf{1}_G$. The classical solution associated with a given $\mflux$ takes the simple form $f = 2\pi\mflux/\A$. We will denote with
\begin{align}\label{EQ:S_cl_undeformed}
    S_{\text{cl}}(\mflux) = \frac{2\pi^2}{g_{\text{YM}}^2\,\A} \, |\mflux|^2
\end{align}
the classical action \eqref{EQ:S_YM_with_f} evaluated on such a configuration.
For generic $\mathbf{g}$, the moduli space of classical solutions has an additional factor of $H^{2\mathbf{g}}$ due to the presence of flat connections wrapping the nontrivial cycles of $\Sigma$.
We will come back to this form of the partition function later in this section.

The physical Hilbert space of quantum states associated with some circle $C\subset\Sigma$ consists of class functions $\Psi(A)$ of the holonomy\footnote{
    While the holonomy depends on the choice of a basepoint $x\in C$, notice that a class function of $U$ does not depend on $x$.
}
\begin{align}
    U = \operatorname{Pexp}\oint_{C,x} (-\mathrm{i}A) \;.
\end{align}
A convenient orthonormal basis for class functions is given by characters of inequivalent irreducible representations of $G$. Therefore, we can always decompose a wavefunction as
\begin{align}
    \Psi(A) = \sum_{R} c_R \, \chi_R(U) \;.
\end{align}
In fact, we can extend \eqref{EQ:Z_YM} to the case where $\Sigma$ has $b$ boundaries \cite{Witten:1991we,Witten:1992xu}. The partition function now carries a dependence on the boundary holonomies $U_1$, $\ldots$, $U_b$:
\begin{align}\label{EQ:Z_YM_boundaries}
    Z(\alpha;U_1,\ldots,U_b)
    &= \sum_{R} (\dim R)^{2-2\mathbf{g}-b} \, e^{-g^2_{\text{YM}}\A\,C_2(R)/2} \, \chi_R(U_1) \ldots \chi_R(U_b) \;.
\end{align}
Surfaces can be glued together similarly to what happens in the context of conventional topological field theories. The associated partition functions are glued together by integrating the holonomy of the common boundary against the Haar measure of $G$ and rely on the orthogonality properties of characters. A flip in the orientation of a boundary corresponds to taking the inverse holonomy $U^{-1}$.

\subsection{The \texorpdfstring{$\mathrm{U}(N)$}{U(N)} theory on the sphere}\label{SEC:review_YM_UN}
In later sections, we will mainly consider the case where $G\simeq\mathrm{U}(N)$, and we will regard the partition function as a function of the rank $N$ and of the effective 't~Hooft coupling $\alpha = g^2_{\text{YM}} N \A$.

An irreducible representation $R$ of $\mathrm{U}(N)$ is labeled by its highest weight vector $\boldsymbol{\lambda}\in\mathbb{Z}^N$ of ordered integers
\begin{align}\label{EQ:weyl_chamber_constraint_lambda}
  \lambda_1 \geq \lambda_2 \geq \ldots \geq \lambda_N \;.
\end{align}
The dimension of $R$ and the eigenvalue of its quadratic Casimir are given by
\begin{align}
  \dim R
  &= \prod_{i<j}\left(1-\frac{\lambda_i-\lambda_j}{i-j}\right) \;, \\
  C_2(R)
  &= \sum_{i=1}^N \lambda_i(\lambda_i-2i+N+1) \;.
\end{align}
For the purpose of applying this to \eqref{EQ:Z_YM}, it is useful to rewrite the above in terms of new variables $\ell_i = -\lambda_i-i+N$. The constraint \eqref{EQ:weyl_chamber_constraint_lambda} restricting to the fundamental Weyl chamber now reads
\begin{align}\label{EQ:weyl_chamber_constraint_ell}
    \ell_1 < \ell_2 < \ldots < \ell_N \;.
\end{align}
With the new variables, the dimension
\begin{align}
  \dim R = \frac{\Delta(\ell_1,\ldots,\ell_N)}{G(N+1)} \;,
\end{align}
is expressed in terms of the Vandermonde determinant, defined as
\begin{align}
  \Delta(\ell_1,\ldots,\ell_N)
  = \det\begin{pmatrix}
    \ell_1^0 & \ell_1^1 & \cdots & \ell_1^{N-1} \\
    \ell_2^0 & \ell_2^1 & \cdots & \ell_2^{N-1} \\
      \vdots &   \vdots & \ddots & \vdots \\
    \ell_N^0 & \ell_N^1 & \cdots & \ell_N^{N-1} \\
  \end{pmatrix}
  = \prod_{i<j}(\ell_j-\ell_i) \;.
\end{align}
Here, $G$ is the Barnes function. The eigenvalue of the quadratic Casimir reads
\begin{align}
    C_2(R) &= \frac{N(1-N^2)}{12} + \smod{\boldsymbol{\ell}} \;,
\end{align}
where we have introduced the shorthand
\begin{align}
    \smod{\boldsymbol{\ell}} = \sum_{i=1}^N \left(\ell_i-\frac{N-1}{2}\right)^2 \;.
\end{align}

Focusing on the sphere topology (i.e.\ on $\mathbf{g}=0$), since both $\dim R$ and $C_2(R)$ are invariant under permutations of the $\ell_i$'s, and since $\dim R$ vanishes whenever two of these coincide, we can simply lift the constraint \eqref{EQ:weyl_chamber_constraint_ell} and normalize appropriately. This gives
\begin{align}\label{EQ:Z_as_sum_over_els}
    Z(\alpha) &= \sum_{\boldsymbol{\ell}\in\mathbb{Z}^N} \xhat{\mathpzc{z}}_{\boldsymbol{\ell}}(\alpha) \cr
    &= \sum_{\boldsymbol{\ell}\in\mathbb{Z}^N} \frac{e^{\alpha(N^2-1)/24}}{N!\,G^2(N+1)} \; \Delta^{2}(\ell_1,\ldots,\ell_N) \; e^{-\frac{\alpha}{2N} \smod{\boldsymbol{\ell}}} \;.
\end{align}

Through the Poisson summation formula, the partition function \eqref{EQ:Z_as_sum_over_els} can be recast in terms of a dual representation
\begin{align}\label{EQ:Z_from_z}
    Z(\alpha) = \sum_{\mflux\in\mathbb{Z}^N} \mathpzc{z}_{\mflux}(\alpha) \;,
\end{align}
where
\begin{align}\label{EQ:z_definition}
    \mathpzc{z}_{\mflux}(\alpha) = \int_{\mathbb{R}^N} \mathrm{d}\ell_1 \ldots \mathrm{d}\ell_N \; e^{-2\pi\mathrm{i}\mflux\cdot\boldsymbol{\ell}} \; \xhat{\mathpzc{z}}_{\boldsymbol{\ell}}(\alpha) \;.
\end{align}
This is nothing but the instanton representation mentioned earlier, where now we regard $\mflux$ as a set of $N$ integers through the natural isomorphism $\Lambda_G \simeq \mathbb{Z}^{\rk G}$.
The physical interpretation as a sum over classical configurations becomes manifest upon performing the Fourier transform above. In fact, one finds that each term in \eqref{EQ:Z_from_z} has the form
\begin{align}\label{z_m_as_w_times_exp}
    \mathpzc{z}_{\mflux}(\alpha) = w_{\mflux}(\alpha) \; e^{-S_{\mathrm{cl}}(\mflux)} \;,
\end{align}
where the function \cite{Minahan:1993tp}
\begin{align}\label{EQ:w_minahan}
    w_{\mflux}(\alpha)
    &= (-1)^m \, \frac{e^{\alpha(N^2-1)/24}}{N!\,G^2(N+1)} \left(\frac{2\pi N}{\alpha}\right)^{\!N^2} \cr
    &\qquad \times \int \mathrm{d}x_1 \ldots \mathrm{d}x_N \; e^{-S_{\mathrm{cl}}(\boldsymbol{x})} \; \prod_{i<j}\left[(x_i-x_j)^2-(\mathfrak{m}_i-\mathfrak{m}_j)^2\right]
\end{align}
captures quantum fluctuation about the classical configuration. In the above, we used the notation
\begin{align}\label{EQ:sigma_definition}
    m = (N-1)\sum_i\mathfrak{m}_i \;.
\end{align}
In particular, the partition function associated with the zero-flux sector reads \cite{Gross:1994mr}
\begin{align}
    \mathpzc{z}_{\boldsymbol{0}}(\alpha) &= C_N \, e^{\alpha(N^2-1)/24} \, \alpha^{-N^2/2} \;, \label{EQ:Z0_undeformed}
\end{align}
where
\begin{align}
    C_N = \frac{(2\pi)^{N/2}N^{N^2/2}}{G(N+1)} \;.
\end{align}

\section{\texorpdfstring{\boldmath $T\bar{T}$}{TT} deformation of Yang--Mills theory}\label{SEC:TT_deformed_YM_theory}
\subsection{The deformed abelian action}
The $T\bar{T}$ deformation of a Lagrangian two-dimensional quantum field theory with action $S$ is described by the flow equation
\begin{align}\label{EQ:flow_equation_action}
    \partial_\mu S = \int_\Sigma \eta \; \TT \;.
\end{align}
The irrelevant operator triggering the deformation reads
\begin{align}\label{EQ:O_TT}
    \TT = T_{\kappa\lambda} T_{\rho\sigma} \epsilon^{\kappa\rho} \epsilon^{\lambda\sigma} \;.
\end{align}

We can apply the above to the case at hand by starting from the action in \eqref{EQ:S_YM_with_f}. We first consider the abelian case. Since $f$ is the only local gauge-invariant scalar degree of freedom of the theory, one can assume that the deformed Lagrangian density will be some function of $f$. In fact, we can equivalently define it as $\mathscr{L}(u,\mu)$, i.e.\ as a function of the deformation parameter $\mu$ and of
\begin{align}
    u = \frac{f^2}{2g^2_{\text{YM}}} \;,
\end{align}
which is the undeformed Lagrangian density itself. With this choice, we have $\mathscr{L}(u,0) = u$. We can compute the stress-tensor of the deformed theory by varying the action with respect to the metric. Since $f$ is defined as the Hodge dual of the field strength, it carries a dependence on the metric and contributes to the variation
\begin{align}
    \delta S_{\text{YM}} 
    &= \delta \int_\Sigma \eta \; \mathscr{L}(u,\mu) \cr
    &= \int_\Sigma \eta \left(\frac{1}{2}\mathscr{L}(u,\mu) - u\,\partial_u\mathscr{L}(u,\mu)\right) g^{\kappa\lambda} \,\delta g_{\kappa\lambda} \;.
\end{align}
From the above, we can easily read off the expression of $T^{\kappa\lambda}$ that, in turn, can be plugged into the flow equation \eqref{EQ:flow_equation_action}. This produces an equation for the Lagrangian density,
\begin{align}\label{EQ:flow_equation_lagrangian}
    \partial_\mu \mathscr{L}
    &= \TT \cr
    &= 2(\mathscr{L} - 2u \, \partial_u\mathscr{L})^2 \;,
\end{align}
that we solve using the ansatz
\begin{align}
    \mathscr{L}(u,\mu) = \sum_{n=0}^\infty \mu^n \mathscr{L}_n(u) \;,
\end{align}
with $\mathscr{L}_0(u) = u$, as mentioned above. We find
\begin{align}
    \mathscr{L}_n(u) = \frac{3\,(4n+1)!}{n!\,(3n+3)!}\,(2u)^{n+1} \;,
\end{align}
which, upon summation, gives
\begin{align}\label{EQ:deformed_lagrangian}
    \mathscr{L} = \frac{3}{8\mu}\left({}_3F_2\mleft(-\frac{1}{2},-\frac{1}{4},\frac{1}{4};\frac{1}{3},\frac{2}{3};\frac{512}{27}\,\mu\,\mathscr{L}_0\mright)-1\right) \;.
\end{align}

We can repeat the analysis for the nonabelian theory. In principle, one is now faced with the choice of which trace structure to include in the deformed action. However, since the undeformed theory, and therefore its stress-tensor, only contain $\tr f^2$, one can safely assume that no other term could appear in the deformed Lagrangian density. With this in mind, we simply redefine
\begin{align}
    u = \frac{\tr f^2}{2g^2_{\text{YM}}} \;,
\end{align}
and repeat the steps above to find that \eqref{EQ:deformed_lagrangian} holds for the nonabelian theory as well.

Notice that \eqref{EQ:deformed_lagrangian} has a branch cut for $\mu\,\mathscr{L}_0>27/512$. This feature is not entirely unexpected, as it appears in other instances of $T\bar{T}$-deformed Lagrangians \cite{Bonelli:2018kik}, but poses a problem if one tries to quantize the deformed theory by starting from \eqref{EQ:deformed_lagrangian}. Since we will take a different route to the quantum theory, we will defer this discussion to Section~\ref{SEC:semiclassical_limit}, which is devoted to the semiclassical limit.

\subsection{The deformed partition function}
We now allow for a general gauge group $G$ and start with an ansatz for a deformed action which is a generalization of \eqref{EQ:S_top},
\begin{align}\label{EQ:S_top_deformed}
    S_{\text{top}} = \mathrm{i}\int_\Sigma \tr(\phi F) + \frac{g^2_{\text{YM}}}{2} \int_\Sigma \eta \, \mathscr{U}(v,\mu) \;,
\end{align}
where $v = \tr \phi^2$. The undeformed Yang--Mills action is recovered with $\mathscr{U}(v,0) = v$.
Again, note that the one defined above is not the most general potential that can be considered, since one could in principle involve other invariant polynomials in $\mathfrak{g}$. However, because of the initial condition, no other term can enter the deformed action.
From the variation
\begin{align}
    \delta S_{\text{top}} = \frac{g^2_{\text{YM}}}{4} \int_\Sigma \eta \; \mathscr{U}(v,\mu) \, g^{\kappa\lambda} \,\delta g_{\kappa\lambda}
\end{align}
one can read off the expression for the stress-energy tensor and plug it in \eqref{EQ:flow_equation_action} to obtain an equation for $\mathscr{U}$,
\begin{align}\label{EQ:U_flow_equation}
    \partial_\mu \mathscr{U}(v,\mu)
    &= 2\TT/g^2_{\text{YM}} \cr
    &= g^2_{\text{YM}}\,\mathscr{U}^2(v,\mu) \;.
\end{align}

Let us now proceed in analogy with \cite{Witten:1992xu} and consider an initial-value circle $C\subset\Sigma$. In a neighborhood of $C$, we write the volume form in terms of local coordinates as $\eta = \mathrm{d}s \wedge \mathrm{d}t$, where $C$ corresponds to $t=0$ and $s$ is a coordinate along $C$ such that $\oint \mathrm{d}s = 1$. Since the action \eqref{EQ:S_top_deformed} is linear in $F$, the Hamiltonian reads
\begin{align}
    H = \frac{g^2_{\text{YM}}}{2} \oint_C\mathrm{d}s \; \mathscr{U}(v,\mu) \;,
\end{align}
and generates translations along $t$. When acting on the representation basis, as in \eqref{EQ:Z_YM_boundaries}, the Hamiltonian is diagonal and takes the simple form
$H = g^2_{\text{YM}}/2 \, \mathscr{U}(C_2(R),\mu)$.

If we now consider a finite cylinder spanned by the range $t\in[0,a]$, the associated partition function will depend on the relevant couplings as $e^{-aH(\mu)}$, where $a$ is the area of the cylinder. As a consequence, one concludes that the deformed partition function obeys the flow equation \cite{Cavaglia:2016oda,Ireland:2019vvj,Santilli:2020qvd}
\begin{align}
    \frac{\partial Z}{\partial \mu} + 2a \, \frac{\partial^2 Z}{\partial a^2} = 0 \;.
\end{align}
We remark that the differential equation above is fully general, since there is nothing special about the chosen topology. In fact, it still applies if we consider, for instance, a disk or a sphere partition function. We simply need to shrink the boundary circles to points, and in doing so, impose trivial holonomies on them. Arbitrary topologies can be further obtained through gluing, thus exploiting the quasi-topological character of the theory.

Before moving on, let us also mention that, while one can safely employ \eqref{EQ:O_TT} for the deformation of the classical action, at the quantum level things are more subtle as one needs to deal with potential ambiguities associated with the UV behavior of composite operators. More precisely, the deformation operator $\TT$ is typically only defined on flat backgrounds where one employs point-split regularization and shows that the pinching limit is actually regular, up to derivative terms. However, in quantum theories described by the action \eqref{EQ:S_top_deformed}, correlators of gauge-invariant local operators are topological, i.e.\ do not depend on the position of the operator insertions. As a consequence, the regularity of the pinching limit of such operators is trivially guaranteed.

We can now specialize to $G\simeq\mathrm{U}(N)$ and use the effective coupling $\alpha$ introduced in Section~\ref{SEC:review_YM_UN}. Moreover, for later convenience we introduce the rescaled deformation parameter $\tau = \mu N^3 g_{\text{YM}}^2$.
In terms of these variables, the flow equation for the partition function reads
\begin{align}\label{EQ:Z_flow_equation}
    \mathbf{F}_{\alpha,\tau} \; Z(\alpha,\tau) = 0 \;,
\end{align}
where we have introduced the differential operator
\begin{align}
    \mathbf{F}_{\alpha,\tau} = \frac{\partial}{\partial\tau} + \frac{2\alpha}{N^2} \frac{\partial^2}{\partial\alpha^2} \;.
\end{align}

It is easy to obtain a closed expression for the deformed Hamiltonian by solving the differential equation for $\mathscr{U}(v,\mu)$, \eqref{EQ:U_flow_equation}:
\begin{align}\label{EQ:deformed_Hamiltonian}
    H = \frac{g^2_{\text{YM}}C_2(R)/2}{1-\tau C_2(R)/N^3} \;.
\end{align}
However, for any $\tau \neq 0$ there is always an infinite number of representations whose energy is arbitrary close to the limit value $-1/(2\mu)$. Consequently, the partition function defined through such a deformed Hamiltonian would necessarily diverge for $\mathbf{g}<2$.

For $\tau>0$, the Hamiltonian, intended as a function of $C_2$, is pathological at $C_2 = N^3/\tau$. Namely, $H\to\mp\infty$ as $C_2\to(N^3/\tau)^\pm$. One would suspect that \eqref{EQ:deformed_Hamiltonian} should really only hold for representations for which $C_2<N^3/\tau$. Extending \eqref{EQ:deformed_Hamiltonian} beyond said range appears devoid of any physical meaning. However, if one hopes to determine the partition function as a sum over representations, one must necessarily understand how the deformation acts on the entire spectrum, not just a portion of it. From a physical standpoint, one is led to postulate that whenever a given representation falls out of the allowed range for the solution \eqref{EQ:deformed_Hamiltonian}, it should be removed from the physical spectrum. Yet, this requirement leads to nontrivial analytic properties for the deformed partition function $Z(\alpha,\tau)$ and makes the study of such a quantity as a solution of the flow equation \eqref{EQ:Z_flow_equation} less obvious.

For $\tau<0$, the situation is more subtle. The partition function is still naively divergent, but the deformed spectrum \eqref{EQ:deformed_Hamiltonian} appears well-defined in any range of values. To obtain a well-defined partition function, one is led to study solutions of \eqref{EQ:Z_flow_equation} that involve instanton-like corrections in the deformation parameter.\footnote{
    The presence of nonperturbative ambiguities in the context of $T\bar{T}$-deformed theories has been studied in \cite{Aharony:2018bad,Iliesiu:2020zld,Griguolo:2021wgy}. See also 
}
However, the question of how to unambiguously determine such nonperturbative corrections is nontrivial and will be addressed in detail in later sections.

For both signs of $\tau$, our approach will be to construct solutions of the flow equation \eqref{EQ:Z_flow_equation} for each flux sector of the theory. In fact, since $\mathbf{F}_{\alpha,\tau}$ is linear, one can construct the full deformed partition function as a sum over individual deformed flux sectors $\mathpzc{z}_{\mflux}(\alpha,\tau)$ obeying $\mathbf{F}_{\alpha,\tau}\,\mathpzc{z}_{\mflux}(\alpha,\tau) = 0$.

\section{\boldmath The zero-flux sector}\label{SEC:z0}
We start by studying the deformed zero-flux sector, describing quantum fluctuations about the ``trivial'' vacuum. In the present section, our approach is to regard $\mathpzc{z}_{\boldsymbol{0}}$ as a power series in the deformation parameter $\tau$.
To this end, it is convenient to introduce the differential operator
\begin{align}\label{EQ:D_operator}
    \mathbf{D}_{\alpha,\tau} = \sum_{n=0}^\infty \frac{\tau^n}{n!} \bigg({-}\frac{2\alpha}{N^2}\frac{\partial^2}{\partial\alpha^2}\bigg)^{\!n} \;.
\end{align}
Since $\mathbf{F}_{\alpha,\tau} \circ \mathbf{D}_{\alpha,\tau} = 0$, $\mathbf{D}_{\alpha,\tau}$ effectively generates power-series solutions of the flow equation when acting on the corresponding ``undeformed'' function encoding the initial condition at $\tau = 0$. However, the sum in \eqref{EQ:D_operator} should be regarded as a formal power series in $\tau$, since in general, it could have vanishing radius of convergence.

Before applying the above to $\mathpzc{z}_{\boldsymbol{0}}$, it is convenient to write the undeformed zero-flux partition function \eqref{EQ:Z0_undeformed} as a power series in $\alpha$,\footnote{To enforce convergence for small $\alpha$, one can regard $N$ as a complex number and choose an appropriate region in the complex $N$-plane.}
\begin{align}\label{EQ:Z0_undeformed_as_series}
    \mathpzc{z}_{\boldsymbol{0}}(\alpha,0) = C_N \sum_{j=0}^\infty \frac{\alpha^{j-N^2/2}}{j!} \left(\frac{N^2-1}{24}\right)^{j} \;.
\end{align}
Then,
\begin{align}\label{Z0_deformed_as_sum}
    \mathpzc{z}_{\boldsymbol{0}}(\alpha,\tau)
    &= \mathbf{D}_{\alpha,\tau} \, \mathpzc{z}_{\boldsymbol{0}}(\alpha,0) \cr
    &= C_N \sum_{j=0}^\infty \frac{\alpha^{j-N^2/2}}{j!} \left(\frac{N^2-1}{24}\right)^{j} \sum_{n=0}^{\infty} \left(\frac{2\tau\vphantom{N^2}}{N^2\alpha}\right)^{\!n} \omega_n \;,
\end{align}
where
\begin{align}
    \omega_n = \frac{(-1)^n\,\Gamma(j-N^2/2)\,\Gamma(1+j-N^2/2)}{n!\,\Gamma(j-n-N^2/2)\,\Gamma(1+j-n-N^2/2)} \;.
\end{align}
Let us now consider the sum over $n$: as anticipated, the series
\begin{align}
    \Phi(t) = \sum_{n=0}^\infty \omega_n \, t^{-n}
\end{align}
is asymptotic, having $\omega_n \sim n!$ for large $n$. In order to apply the standard machinery of Borel resummation, we first consider its Borel transform,
\begin{align}\label{EQ:Borel_transform_Z0}
    \mathcal{B}\Phi(\zeta)
    &= \sum_{n=0}^{\infty} \zeta^{n} \frac{\omega_n}{n!} \cr
    &= {}_2F_1(N^2/2-j,N^2/2-j+1;1;-\zeta) \;.
\end{align}
We observe two different behaviors depending on sign of $\tau$.

\subsection{\texorpdfstring{$\tau>0$}{tau > 0}}
For positive values of $\tau$, and hence of
\begin{align}
    t = \frac{N^2\alpha}{2\tau\vphantom{N^2}} \;,
\end{align}
we can simply Borel-resum the above by taking a Laplace transform along the positive real axis in the complex $\zeta$-plane
\begin{align}
    \mathcal{S}_0\Phi(t)
    &= t\int_0^\infty \mathrm{d} \zeta \; \Phi(\zeta) \, e^{-t\zeta} \cr
    &= t^{N^2/2-j} \; U(N^2/2-j,0,t) \;.
\end{align}
Plugging this expression back in \eqref{Z0_deformed_as_sum} gives
\begin{align}
    \mathpzc{z}_{\boldsymbol{0}}(\alpha,\tau)
    &= C_N \left(\frac{N^2}{2\tau}\right)^{\!N^2/2} \sum_{j=0}^\infty \frac{1}{j!} \left(\frac{\tau(N^2-1)}{12N^2}\right)^{j} U\mleft(\frac{N^2}{2}-j,0,\frac{N^2\alpha}{2\tau}\mright) \;.
\end{align}

The sum is easily performed through the multiplication theorem for the Tricomi confluent hypergeometric function. This leads to the final expression
\begin{align}\label{Z0_tau_positive}
    \mathpzc{z}_{\boldsymbol{0}}(\alpha,\tau)
    &= C_N \, e^{\X} \, \Y^{N^2/2} \, U(N^2/2,0,\W) \;,
\end{align}
where we defined
\begin{align}
    \X &= \frac{N^2(N^2-1)\alpha}{2(N^2(12+\tau)-\tau)} \;, \\
    \Y &= \frac{N^2(12+\tau)-\tau}{24\tau} \;, \\
    \W &= \frac{6N^4\alpha}{\tau(N^2(12+\tau)-\tau)} \;.
\end{align}
It is immediate to check that, indeed, \eqref{Z0_tau_positive} is a solution of the flow equation \eqref{EQ:Z_flow_equation} and reproduces the correct undeformed limit \eqref{EQ:Z0_undeformed} for $\tau\to 0^+$.

\subsection{\texorpdfstring{$\tau<0$}{tau < 0}}
When $\tau<0$, which means $t<0$, the series should be resummed by taking a directional Laplace transform along the negative real $\zeta$ axis. However, the Borel transform \eqref{EQ:Borel_transform_Z0} has a Stokes line at $\arg \zeta = \pi$ due to a cut that extends over $\zeta\in(-\infty,-1)$. By approaching the cut from above and below with lateral Laplace transforms one finds
\begin{align}
    \mathcal{S}_{\pm \pi}\Phi(t)
    &= t\int_0^{e^{\pm\mathrm{i}\pi}\infty} \mathrm{d} \zeta \; e^{-t\zeta} \, {}_2F_1(N^2/2-j,N^2/2-j+1;1;-\zeta) \cr
    &= -t\int_0^{\infty} \mathrm{d} x \; e^{tx} \, {}_2F_1(N^2/2-j,N^2/2-j+1;1;-e^{\pm\mathrm{i}\pi}x) \cr
    &= (t\mp\mathrm{i}0)^{N^2/2-j} \; U(N^2/2-j,0,t\mp\mathrm{i}0) \;. \vphantom{\int^\infty}
\end{align}
\begin{figure}[tbp]
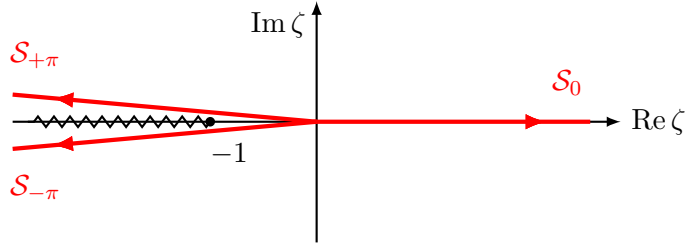

    \centering\lateralborelym
    \caption{\label{FIG:lateral_borel} The contours where the directional Laplace transformations are taken. For $\tau>0$, $\mathcal{S}_0$ gives the full result of the Borel summation, free from nonperturbative ambiguities. The Stokes line on the negative $\zeta$ axis induces two different lateral Laplace transformations, $\mathcal{S}_{+\pi}$ and $\mathcal{S}_{-\pi}$, that are relevant for the $\tau<0$ regime.}
\end{figure}
We can use the analytic properties of the Tricomi confluent hypergeometric function to rewrite the above as
\begin{align}
    \mathcal{S}_{+\pi}\Phi(t)
    &= (t-\mathrm{i}0)^{N^2/2-j} \left[U(N^2/2-j,0,t) - \frac{2\mathrm{i}\pi t}{\Gamma(N^2/2-j)} \, {}_1F_1(N^2/2-j+1;2;t) \right] \;, \cr
    \mathcal{S}_{-\pi}\Phi(t)
    &= (t+\mathrm{i}0)^{N^2/2-j} \; U(N^2/2-j,0,t) \;.
\end{align}
Since we are interested in a real partition function for real values of the deformation parameter $\tau$ and the effective 't Hooft coupling $\alpha$, we can remove the nonperturbative ambiguities associated with the presence of the Stokes line by employing a prescription known as \emph{median resummation},
\begin{align}
    \mathcal{S}_{\mathrm{med}}\Phi(t) = \frac{1}{2}(\mathcal{S}_{+\pi}\Phi(t) + \mathcal{S}_{-\pi}\Phi(t)) \;.
\end{align}

For odd $N$, this prescription leads to
\begin{align}\label{EQ:Z_tau_negative_N_odd}
    \mathpzc{z}_{\boldsymbol{0}}(\alpha,\tau)
    &= \frac{\pi C_N}{\Gamma(N^2/2)} \, \alpha \left(-\frac{N^2}{2\tau}\right)^{\!N^2/2+1} \cr
    &\qquad \times \sum_{j=0}^\infty \frac{(1-N^2/2)_j}{j!} \left(-\frac{N^2-1}{N^2}\frac{\tau}{12}\right)^{j} \, {}_1F_1\left(\frac{N^2}{2}-j+1;2;\frac{N^2\alpha}{2\tau}\right) \cr
    &= - \frac{\pi C_N}{\Gamma(N^2/2)} \, \W e^{\X} (-\Y)^{N^2/2} \, {}_1F_1(N^2/2+1;2;\W) \;.
\end{align}
For $N$ even, we find instead
\begin{align}\label{EQ:Z_tau_negative_N_even}
    \mathpzc{z}_{\boldsymbol{0}}(\alpha,\tau)
    &= C_N \left(-\frac{N^2}{2\tau}\right)^{\!N^2/2} \sum_{j=0}^\infty \frac{1}{j!} \left(\frac{N^2-1}{N^2}\frac{\tau}{12}\right)^{j} \cr
    &\qquad\times \left[U\mleft(\frac{N^2}{2}-j,0,\frac{N^2\alpha}{2\tau}\mright) - \frac{\mathrm{i}\pi}{\Gamma(N^2/2-j)} \frac{N^2\alpha}{2\tau} \, {}_1F_1\left(\frac{N^2}{2}-j+1;2;\frac{N^2\alpha}{2\tau}\right) \right] \cr
    &= C_N \, e^{\X} \, \Y^{N^2/2} \left[ U(N^2/2,0,\W) - \frac{\mathrm{i}\pi\,\W}{\Gamma(N^2/2)} \, {}_1F_1(N^2/2+1;2;\W) \right] \;.
\end{align}
Because the second term is purely imaginary, it necessarily cancels the imaginary part of the first one. In fact, we can rewrite the two expressions as a single formula that holds for any $N$ as
\begin{align}
    \mathpzc{z}_{\boldsymbol{0}}(\alpha,\tau)
    &= \operatorname{Re}\mleft(C_N \, e^{\X} \, \Y^{N^2/2} \, U(N^2/2,0,\W)\mright) \;.
\end{align}

In both cases, we made use of multiplication theorems \eqref{EQ:Tricomi_multiplication_theorem} and \eqref{EQ:Kummer_multiplication_theorem} for confluent hypergeometric functions. For the theorems to hold, we need $\tau>\tau_{\text{min}}$, where
\begin{align}
    \tau_{\text{min}} = -\frac{12N^2}{N^2-1} \;.
\end{align}
In fact, when approaching $\tau_{\text{min}}$ from the left, $\W\to+\infty$. Consequently, the partition function diverges in this limit since the instanton-like terms blow up. We are therefore forced to regard $\tau>\tau_{\text{min}}$ as a constraint on the validity of \eqref{EQ:Z_tau_negative_N_odd} and \eqref{EQ:Z_tau_negative_N_even}. When this condition is obeyed, the two expressions above are real, satisfy the flow equation, and reproduce the undeformed limit for $\tau \to 0$. One could, in principle, extend the range of validity by studying the relevant nonperturbative contributions for $\tau<\tau_{\text{min}}$. However, upon summing over $\mflux$, we will find an expression for the total partition function that does not exhibit any pathological behavior at $\tau = \tau_{\text{min}}$ and that can be taken to hold for any value of the deformation parameter.

\section{Any flux sector}\label{SEC:zm}
The results of the previous section suggest an ansatz for the structure of the full solution of the flow equation \eqref{EQ:Z_flow_equation} when written in terms of the variables $\Y$ and $\W$,\footnote{Here, $\X$ should be regarded as the shorthand
\begin{align*}
    \X = \frac{(N^2-1)\W}{1+24\Y-N^2} \;.
\end{align*}}
\begin{align}
    \mathpzc{z}_{\mflux}(\alpha,\tau) = C_N \, e^{\X} \, \Y^{N^2/2} \, f(\Y,\W) \;.
\end{align}
On the above, the flow equation becomes
\begin{align}
    \W \partial_{\W} f(\Y,\W) - \W \partial^2_{\W} f(\Y,\W) + \Y \partial_{\Y} f(\Y,\W) + \frac{N^2}{2} f(\Y,\W) = 0 \;.
\end{align}
This equation can be easily solved by separation of variables. Specifically, if we choose $f(\Y,\W) = \Upsilon(\Y)\,\Omega(\W)$, we obtain two ordinary equations
\begin{align}
    \Y \Upsilon'(\Y) &= s \Upsilon(\Y) \;, \\
    \W \Omega''(\W) - \W \Omega'(\W) - (N^2/2+s)\,\Omega(\W) &= 0 \;,
\end{align}
with solutions
\begin{align}
    \Upsilon(\Y) &= c \, \Y^s \;, \\
    \Omega(\W) &= u \; U(N^2/2+s,0,\W) + v \, \W \; {}_1F_1(N^2/2+s+1;2;\W) \;.
\end{align}
Here, $s$ is just an integration constant labeling different solutions.

We rewrite the ansatz as a generic linear combination of the above, where the coefficients are chosen in order to reproduce the boundary value at $\tau = 0$, which is fixed by the undeformed theory. From \eqref{z_m_as_w_times_exp} and \eqref{EQ:w_minahan}, it is easy to see that a generic undeformed flux sector can be expressed as a convergent expansion in $1/\alpha$ with structure
\begin{align}\label{EQ:z_m_undeformed_ansatz}
    \mathpzc{z}_{\mflux}(\alpha,0) = \sum_{k=0}^{\infty} \frac{a_{\mflux,k}}{\alpha^{k+N^2/2}} \;.
\end{align}

The behavior of $\mathpzc{z}_{\mflux}(\alpha,\tau)$ for small $\tau$ is sensitive to the sign of the deformation. We separately study the two choices.

\subsection{\texorpdfstring{$\tau>0$}{tau > 0}}
When $\tau\to0^+$, $\W\to+\infty$ and ${}_1F_1\mleft(N^2/2+s+1,2,\W\mright)$ becomes exponentially divergent. This is not surprising as one expects the Kummer confluent hypergeometric function to bring nonperturbative contributions in $\tau$, which should be absent for positive values of the deformation parameter. Therefore, when $\tau>0$, we consider a general solution of the flow equation written as the linear combination
\begin{align}\label{EQ:Z_n_positive_tau}
    \mathpzc{z}_{\mflux}(\alpha,\tau) = C_N \, e^{\X} \, \Y^{N^2/2} \sum_{s=0}^{\infty} \frac{p_{\mflux,s}}{s!} \, (-\Y)^s \; U(N^2/2+s,0,\W) \;.
\end{align}
The undeformed limit is given by
\begin{align}
    \lim_{\tau\to0^+} \mathpzc{z}_{\mflux}(\alpha,\tau) &= \mathpzc{z}_{\boldsymbol{0}}(\alpha,0) \, \sum_{s=0}^{\infty} \frac{p_{\mflux,s}}{s!} \, (-\alpha)^{-s} \;.
\end{align}

For the zero-flux sector, one can trivially determine the coefficients in the sum by comparing with \eqref{EQ:Z0_undeformed}. This gives $p_{\boldsymbol{0},s} = \delta_{s,0}$, which reproduces the result \eqref{Z0_tau_positive} derived in the previous section.
For a generic flux sector, we can determine the coefficients $p_{\mflux,s}$ by exploiting the so-called \emph{Ramanujan's Master Theorem}.\footnote{
    According to the theorem, if a complex function $f$ has an expansion of the form
    $$
        f(x) = \sum_{k=0}^\infty \frac{\varphi(k)}{k!} (-x)^k \;,
    $$
    then its Mellin transform reads
    $$
        \int_0^\infty \mathrm{d}x \; x^{s-1} \, f(x) = \Gamma(s) \, \varphi(-s) \;.
    $$
} We find
\begin{align}\label{EQ:ps_ramanujan}
    p_{\mflux,s} = \frac{1}{\Gamma(-s)} \int_0^\infty \mathrm{d}\alpha \; \alpha^{s-1} \, \frac{\mathpzc{z}_{\mflux}(\alpha,0)}{\mathpzc{z}_{\boldsymbol{0}}(\alpha,0)} \;.
\end{align}
The formula is understood for some region in the complex $s$-plane where the above is well-defined. The result is then analytically continued to positive integer values of $s$.

To determine the coefficient $p_{\mflux,s}$ for a generic flux $\mflux$ we start with the definition \eqref{EQ:Z_from_z}. Performing the integral is nontrivial due to the presence of the square of the Vandermonde determinant in \eqref{EQ:Z_as_sum_over_els}. However, within the Fourier integral, we can trade it for the differential operator
\begin{align}\label{EQ:differential_operator_V}
    \V = \frac{\Delta^2(\partial_{\mathfrak{m}_1},\ldots,\partial_{\mathfrak{m}_N})}{(-4\pi^2)^{N(N-1)/2}} \;,
\end{align}
leading to
\begin{align}
    \mathpzc{z}_{\mflux}(\alpha,0)
    &= \mathpzc{z}_{\boldsymbol{0}}(\alpha,0) \int_{\mathbb{R}^N} \mathrm{d}\ell_1 \ldots \mathrm{d}\ell_N \; e^{-2\pi\mathrm{i}\mflux\cdot\boldsymbol{\ell}} \; \frac{\Delta^{2}(\ell_1,\ldots,\ell_N) \; e^{-\frac{\alpha}{2N} \smod{\boldsymbol{\ell}}}}{(2\pi)^{N/2} \, N! \, G(N+1) \, (N/\alpha)^{N^2/2}} \cr
    &= \mathpzc{z}_{\boldsymbol{0}}(\alpha,0) \, \frac{(\alpha/N)^\nu}{N!\,G(N+1)} \, (-1)^m \, \V \,e^{-2\pi^2N|\mflux|^2/\alpha} \;.
\end{align}
We have used the fact that $\V (e^{-\pi\mathrm{i}m}f(\mflux)) = e^{-\pi\mathrm{i}m}\,\V f(\mflux)$.
Now we can plug the above in \eqref{EQ:ps_ramanujan}. The integration can be performed by assuming $\operatorname{Re}s<0$. This gives the coefficient
\begin{align}\label{EQ:p_s_generic}
    p_{\mflux,s}
    &= \frac{(-1)^{m+\nu}N^s}{N!\,G(N+1)} \, \frac{\Gamma(s+1)}{\Gamma(s+\nu+1)} \; \V \, \big(2\pi^2|\mflux|^2\big)^{s+\nu} \;,
\end{align}
written in terms of the operator in \eqref{EQ:differential_operator_V} and the shorthand $\nu = N(N-1)/2$.

\subsection{\texorpdfstring{$\tau<0$}{tau < 0}}\label{SEC:any_zm_negative}
Let us now consider the case where $\tau<0$. From the results of the previous section and the abelian case \cite{Griguolo:2022xcj}, we know that the deformed flux sector should receive nonperturbative corrections in $\tau$. Indeed, the presence of the Kummer confluent hypergeometric function, necessary to construct real solutions of the flow equation \eqref{EQ:Z_flow_equation} at $\tau<0$, brings instanton-like contributions for $\tau \to 0^-$ (see Eq.\ \eqref{EQ:Kummer_asymptotics}). As noticed in the previous section, the results will differ according to the parity of $N$. To avoid repeating the analysis for both choices, the rest of the present work will only focus on the case where $N$ is odd, whenever $\tau<0$. Upon taking the sum over all flux sectors, we will find an expression for the full deformed partition function that applies to any $N$. 

We choose the ansatz
\begin{align}\label{EQ:Z_n_negative_tau_odd_N}
    \mathpzc{z}_{\mflux}(\alpha,\tau) = - \pi C_N \, \W e^{\X} (-\Y)^{N^2/2} \sum_{s\in K} \frac{(-1)^{2s} \, p_{\mflux,s}}{s!\,\Gamma(s+N^2/2)} \, (-\Y)^s \; {}_1F_1(N^2/2+s+1;2;\W) \;.
\end{align}
Let us for the moment set $K = \{0,1,2,\ldots\}$. The ansatz is carefully defined in such a way to reproduce the same limit as above, but now taken from negative values of $\tau$, i.e.\ with this choice, it follows from \eqref{EQ:Kummer_asymptotics}
\begin{align}\label{EQ:Z_n_negative_tau_undeformed_limit}
    \lim_{\tau\to0^-} \mathpzc{z}_{\mflux}(\alpha,\tau) &= \mathpzc{z}_{\boldsymbol{0}}(\alpha,0) \, \sum_{s=0}^{\infty} \frac{p_{\mflux,s}}{s!} \, (-\alpha)^{-s} \;,
\end{align}
and one can still use Eq.~\eqref{EQ:ps_ramanujan} to find the coefficients of the sum. This is consistent with the result for the zero-flux sector derived in \eqref{EQ:Z_tau_negative_N_odd}.

However, it is possible to add to the sum in \eqref{EQ:Z_n_negative_tau_odd_N} any half-integer $s$ with $s>1-N^2/2$ without modifying the undeformed limit \eqref{EQ:Z_n_negative_tau_undeformed_limit}. This is due to the fact that for such values of $s$, the Kummer confluent hypergeometric function acts as a purely nonperturbative contribution to the result. As we will see later, the inclusion of these additional terms is crucial to ensure the convergence of the sum over $\mflux$ producing the full partition function $Z(\alpha,\tau)$ and to generate the expected semiclassical limit of each flux sector. We will address both of these important points in the following sections.

For the moment, we simply choose $K = K^+ \cup K^-$, where
\begin{align}
    K^+ &= \bigg\{0,\frac{1}{2},1,\frac{3}{2},2,\ldots\bigg\} \;, \label{EQ:Kplus} \\
    K^- &= \bigg\{1-\frac{N^2}{2},2-\frac{N^2}{2},\ldots,-\frac{1}{2}\bigg\} \;. \label{EQ:Kminus}
\end{align}
As observed at the end of the last section, the ansatz in \eqref{EQ:Z_n_negative_tau_odd_N} holds for $\tau > \tau_{\text{min}}$, since the expression diverges for $\tau \to \tau^-_{\text{min}}$. As we will see in the following, this is a constraint that applies only to the individual flux sectors since the expression for the full deformed partition function will hold for any value of $\tau$.

\section{The full partition function}\label{SEC:Z_full}
In this section, we will compute the full deformed partition function by summing over all the deformed flux sectors. For these, we rely on the results of Section~\ref{SEC:z0} and Section~\ref{SEC:zm}. 

\subsection{\texorpdfstring{$\tau>0$}{tau > 0}}\label{SEC:Z_tau_positive}
For positive values of the deformation parameter, we can use \eqref{EQ:Z_n_positive_tau} to compute a generic deformed flux sector, starting from \eqref{EQ:p_s_generic}. The sum over $s$ is performed by first replacing the Tricomi confluent hypergeometric function with its integral representation \eqref{EQ:Tricomi_integral_representation} and then by using the identity
\begin{align}
    \sum_{s=0}^{\infty} \frac{x^s}{\Gamma(s+a)\,\Gamma(s+b)}
    &= \frac{{}_1F_2(1;a,b;x)}{\Gamma(a)\,\Gamma(b)} \cr
    &= {}_1\tilde{F}_2(1;a,b;x) \;.
\end{align}
This readily gives 
\begin{align}
    \mathpzc{z}_{\mflux}(\alpha,\tau)
    &= \frac{(-1)^{m+\nu} \, C_N \, e^{\X} \, \Y^{N^2/2}}{N!\,G(N+1)} \, \int_0^\infty \mathrm{d}t \; \frac{e^{-t\W}}{t(1+t)} \left(\frac{t}{1+t}\right)^{\!N^2/2} \cr
    &\quad \times \V \, \bigg[\big(2\pi^2|\mflux|^2\big)^{\nu} \, {}_1\tilde{F}_2\Big(1;\tfrac{N^2}{2},\nu+1;-2\pi^2N\Y\tfrac{t}{1+t}|\mflux|^2\Big) \bigg] \;. \cr
\end{align}
The above expression can be simplified by changing integration variable with
\begin{align}
    t &= \frac{r^2}{2N\Y-r^2}
\end{align}
and by taking advantage of the following property of the hypergeometric function:
\begin{align}\label{EQ:1F2_identity_1}
    {}_1\tilde{F}_2\Big(1;\tfrac{N^2}{2},\nu+1;-z^2\Big) = (-1)^\nu z^{1+N/2-N^2} J_{N/2-1}(2z) - \sum_{s=0}^{\nu-1} \frac{(-z^2)^{s-\nu}}{s!\,\Gamma(s+N/2)} \;.
\end{align}
We observe that, when inserted into to our integral, the finite sum appearing above combines with $|\mflux|^{2\nu}$ to produce a polynomial of degree $2(\nu-1)$ in the $\mflux$'s, and it vanishes under the action of $\V$, which is a differential operator of order $2\nu$. We are then left with
\begin{align}\label{EQ:zm_pos_integral_formula}
    \mathpzc{z}_{\mflux}(\alpha,\tau)
    &= \frac{(-1)^{m}\,e^{\X}}{N!\,G^2(N+1)} \, \V \, \Bigg[2\pi\int_0^{\sqrt{2N\Y}} \mathrm{d}r \; r^{N/2} \, e^{-\frac{r^2\W}{2N\Y-r^2}} \, |\mflux|^{1-N/2} \, J_{N/2-1}(2\pi r|\mflux|) \Bigg] \;.
\end{align}

In the above, we recognize the $N$-dimensional Fourier transform\footnote{
    Let $f(\mathbf{x})$ be a spherically-symmetric function on $\mathbb{R}^N$. We denote $f(\mathbf{x}) = F(|\mathbf{x}|)$. Then
    $$
        \int_{\mathbb{R}^N} \mathrm{d}\mathbf{x} \; e^{-2\pi\mathrm{i}\mathbf{k}\cdot\mathbf{x}} \, f(\mathbf{x}) = 2\pi |\mathbf{k}|^{1-N/2} \int_0^\infty \mathrm{d}r \; J_{N/2-1}(2\pi|\mathbf{k}|r) \, r^{N/2} \, F(r) \;.
    $$
}
\begin{align}
    \mathpzc{z}_{\mflux}(\alpha,\tau)
    &= \frac{(-1)^{m}\,e^{\X}}{N!\,G^2(N+1)} \, \V \, \Bigg[\int \mathrm{d}\ell_1\ldots\mathrm{d}\ell_N \; e^{-2\pi\mathrm{i}\mflux\cdot\boldsymbol{\ell}} \; e^{-\frac{\W|\boldsymbol{\ell}|^2}{2N\Y-|\boldsymbol{\ell}|^2}} \; \Theta(2N\Y-|\boldsymbol{\ell}|^2) \Bigg] \cr
    &= \int \mathrm{d}\ell_1\ldots\mathrm{d}\ell_N \; e^{-2\pi\mathrm{i}\mflux\cdot\boldsymbol{\ell}} \; \xhat{\mathpzc{z}}_{\boldsymbol{\ell}}(\alpha,\tau) \;,
\end{align}
of the spherically-symmetric smooth function with compact support,
\begin{align}
    \xhat{\mathpzc{z}}_{\boldsymbol{\ell}}(\alpha,\tau)
    &= \frac{\Theta(2N\Y-\langle\boldsymbol{\ell}\rangle)}{N!\,G^2(N+1)} \; \Delta^2(\ell_1,\ldots,\ell_N) \; e^{\X-\frac{\W\langle\boldsymbol{\ell}\rangle}{2N\Y-\langle\boldsymbol{\ell}\rangle}} \;.
\end{align}
Here, $\Theta$ denotes the Heaviside step function.
The full partition function comes from taking the sum over $\mflux$, which can be traded for a sum over $\boldsymbol{\ell}$ through the Poisson summation formula, 
\begin{align}
    Z(\alpha,\tau)
    &= \sum_{\mflux\in\mathbb{Z}^N} \mathpzc{z}_{\mflux}(\alpha,\tau) \cr
    &= \sum_{\boldsymbol{\ell}\in\mathbb{Z}^N} \xhat{\mathpzc{z}}_{\boldsymbol{\ell}}(\alpha,\tau) \;.
\end{align}

Analogously to the abelian case \cite{Griguolo:2022xcj}, the full partition function can be expressed in terms of the deformed Hamiltonian \eqref{EQ:deformed_Hamiltonian},
\begin{align}\label{EQ:Z_tau_positive}
    Z(\alpha, \tau)
    &= \sum_{R \in \mathcal{R}_{N,\tau}} (\dim R)^{2} \; e^{-\frac{\alpha\vphantom{C_2}}{2N^{\vphantom{3}}} \frac{\phantom{1-\tau\,}C_2(R)\phantom{/N^3}}{1-\tau\,C_2(R)/N^3}} \;.
\end{align}
Crucially, the range of the sum extends over $\mathcal{R}_{N,\tau}$, the set of inequivalent irreducible representations of $U(N)$ with $\tau C_2 < N^3$. As a consequence, for any $\tau>0$ the deformed partition function is a sum over a finite set, which is necessarily convergent. Moreover, the number of terms in the sum in \eqref{EQ:Z_tau_positive} varies with $\tau$. Specifically, as $\tau$ increases, a given representation $R$ drops out of the sum when $\tau$ reaches the critical value $\tau_R = N^3/C_2(R)$. This prevents the partition function from being analytic in $\tau$ at $\tau = \tau_R$. However, as $\tau\to\tau_R^-$, the term in the sum associated with $R$ vanishes together with its derivatives of any order, thus making the partition function smooth. These critical values of $\tau$ have been studied in \cite{Griguolo:2022xcj} for the abelian theory. They were observed to be associated with quantum phase transition of infinite order.

For $\tau > N^2$, only the trivial representation contributes to the sum and the partition function becomes itself trivial with $Z(\alpha,\tau) = 1$.

\subsection{\texorpdfstring{$\tau<0$}{tau < 0}}\label{SEC:Z_tau_negative}
Let us now turn to the case where the deformation parameter is negative and $N$ is odd. We start from \eqref{EQ:Z_n_negative_tau_odd_N} and write $\mathpzc{z}_{\mflux} = \mathpzc{z}^+_{\mflux} + \mathpzc{z}^-_{\mflux}$ by splitting the range of the sum over $K^+$ and $K^-$ respectively, as defined in \eqref{EQ:Kplus} and \eqref{EQ:Kminus}.

In Appendix~\ref{APP:identity} we show that
\begin{align}\label{EQ:zplus_identity}
    \mathpzc{z}^+_{\mflux}(\alpha,\tau)
    &= \frac{e^{\X}}{N!\,G^2(N+1)} \, \V \, \Bigg[2\pi\int_0^{\infty} \mathrm{d}r \; r^{N/2} \, |\mflux|^{1-N/2} \, J_{N/2-1}(2\pi r|\mflux|) \cr
    &\quad \times \bigg(e^{-\frac{r^2\W}{2N\Y-r^2}} - e^{\W} - \W e^{\W}\sum_{k=1}^{k_{\text{max}}}\frac{1}{k}\left(\frac{2N\Y}{r^2}\right)^{\!k}\,L_{k-1}^1(-\W)\bigg) \Bigg] \cr
    &= \int \mathrm{d}\ell_1\ldots\mathrm{d}\ell_N \; e^{-2\pi\mathrm{i}\mflux\cdot\boldsymbol{\ell}} \; \xhat{\mathpzc{z}}^{\!+}_{\boldsymbol{\ell}}(\alpha,\tau) \;,
\end{align}
where $k_{\text{max}} = (N^2-1)/2$, and
\begin{align}\label{EQ:zhatplus}
    \xhat{\mathpzc{z}}^{\!+}_{\boldsymbol{\ell}}(\alpha,\tau)
    &= \frac{e^{\X} \, \Delta^2(\ell_1,\ldots,\ell_N)}{N!\,G^2(N+1)} \; \bigg(e^{-\frac{\W|\boldsymbol{\ell}|^2}{2N\Y-|\boldsymbol{\ell}|^2}} - e^{\W} - \W e^{\W}\sum_{k=1}^{k_{\text{max}}}\frac{1}{k}\left(\frac{2N\Y}{|\boldsymbol{\ell}|^2}\right)^{\!k}\,L_{k-1}^1(-\W)\bigg) \;.
\end{align}

One would be tempted to apply the standard Poisson summation formula to the above. However, the sum over $\mflux$ does not converge.\footnote{
    We prove in \eqref{EQ:z_m_integral_formula} that the sum of $\mathpzc{z}_{\mflux}$ over all flux sectors is convergent. This can be split as
    $$
        \sum_{\mflux\in\mathbb{Z}^N} \mathpzc{z}^+_{\mflux}(\alpha,\tau) + \sum_{\mflux\in\mathbb{Z}^N} \mathpzc{z}^-_{\mflux}(\alpha,\tau) \;.
    $$
    The second sum diverges since $\mathpzc{z}^-_{\mflux}$ is a polynomial in $\mflux$. As a consequence, the first sum must diverge as well, and it must do so by sharing the same behavior at large $\mflux$.
}
One can check that, as a consequence, $\xhat{\mathpzc{z}}^{\!+}_{\boldsymbol{\ell}}(\alpha,\tau)$ diverges for $\boldsymbol{\ell} \to \boldsymbol{0}$. It is possible to subtract the $\boldsymbol{\ell} = \boldsymbol{0}$ term (see e.g.~\cite{DURAN1998581}) with\footnote{\label{FN:cesaro_limit}
In \eqref{EQ:generalized_poisson} and in other instances throughout the remainder of the present section, we will deal with finite expressions written as the difference between a divergent sum and a divergent integral. In such cases, convergence is ensured by taking the appropriate simultaneous limit of the spherical partial sums. Namely, the combination
$$
    \sum_{\mathbf{x}} f(\mathbf{x}) - \int \mathrm{d}\mathbf{x} \; g(\mathbf{x})
$$
should be understood as the regulated form
$$
    \lim_{\Lambda\to\infty}\Bigg(\sum_{|\mathbf{x}|\leq\Lambda} f(\mathbf{x}) - \int_{\raisebox{-2.5pt}{\tiny$|\mathbf{x}|{\leq}\Lambda$}} \!\! \mathrm{d}\mathbf{x} \; g(\mathbf{x}) \Bigg) \;.
$$
}
\begin{align}\label{EQ:generalized_poisson}
    \sum_{\mflux\in\mathbb{Z}^N} \mathpzc{z}^+_{\mflux}(\alpha,\tau) - \int_{\mathbb{R}^N} \mathrm{d}\mflux \; \mathpzc{z}^+_{\mflux}(\alpha,\tau)
    &= \sum_{\boldsymbol{\ell}\neq\boldsymbol{0}} \xhat{\mathpzc{z}}^{\!+}_{\boldsymbol{\ell}}(\alpha,\tau) \;.
\end{align}
We use the above to write an expression for the full deformed partition function,
\begin{align}\label{EQ:Z_neg_final}
    Z(\alpha,\tau) = \sum_{\boldsymbol{\ell}\neq\boldsymbol{0}} \xhat{\mathpzc{z}}^{\!+}_{\boldsymbol{\ell}}(\alpha,\tau) + \mathcal{R}(\alpha,\tau) \;,
\end{align}
where the residual term reads\footnote{We omit the $\mflux=\boldsymbol{0}$ term in the sum, since this vanishes. The result for the zero-flux sector is in \eqref{EQ:Z_tau_negative_N_odd}.}
\begin{align}
    \mathcal{R}(\alpha,\tau)
    &= \int_{\mathbb{R}^N} \mathrm{d}\mflux \; \mathpzc{z}^+_{\mflux}(\alpha,\tau) + \sum_{\mflux\neq\boldsymbol{0}} \mathpzc{z}^-_{\mflux}(\alpha,\tau) \;.
\end{align}

We now have to efficiently express the residual part in terms of $\mathrm{U}(N)$-representation data. To this end, we split the residual term as
\begin{align}
    \mathcal{R}(\alpha,\tau) = \mathcal{R}_{\mathrm{EM}}(\alpha,\tau) + \mathcal{R}_0(\alpha,\tau)
\end{align}
where
\begin{align}
    \label{EQ:calR_EM}
    \mathcal{R}_{\mathrm{EM}}(\alpha,\tau)
    &= \sum_{\mflux\neq\boldsymbol{0}} \mathpzc{z}^-_{\mflux}(\alpha,\tau) - \int_{\mathbb{R}^N} \mathrm{d}\mflux \; \mathpzc{z}^-_{\mflux}(\alpha,\tau) \;, \\
    \label{EQ:calR_0}
    \mathcal{R}_0(\alpha,\tau)
    &= \int_{\mathbb{R}^N} \mathrm{d}\mflux \; \mathpzc{z}_{\mflux}(\alpha,\tau) \;.
\end{align}
We remind the reader that convergence in the definition \eqref{EQ:calR_EM} should be understood as in Footnote~\ref{FN:cesaro_limit}.

From \eqref{EQ:Z_n_negative_tau_odd_N} and \eqref{EQ:p_s_generic}, and by using
\begin{align}
    {}_1F_1(k+1;2;\W) = \frac{e^\W}{k} L_{k-1}^1(-\W) \;,
\end{align}
we find
\begin{align}
    \mathpzc{z}^-_{\mflux}(\alpha,\tau)
    &= \frac{(-1)^{\nu} \pi^{1-N/2} \, \W e^{\X+\W}}{N!\,G^2(N+1)} \sum_{k=1}^{k_{\text{max}}} \frac{(-2\pi^2N\Y)^{k} \; L_{k-1}^1(-\W)}{\Gamma(k+1)\,\Gamma(k+1-N/2)} \; \V \, |\mflux|^{2k-N} \;,
\end{align}
while its Fourier anti-transform reads
\begin{align}
    \xhat{\mathpzc{z}}^-_{\boldsymbol{\ell}}(\alpha,\tau)
    &= \frac{\W e^{\X+\W}}{N!\,G^2(N+1)} \sum_{k=1}^{k_{\text{max}}} \frac{(2N\Y)^{k}}{k} \; L_{k-1}^1(-\W) \; \frac{\Delta^2(\ell_1,\ldots,\ell_N)}{|\boldsymbol{\ell}|^{2k}} \;.
\end{align}
For the last step, we use again a generalized Poisson summation formula \cite{DURAN1998581} to rewrite the residual term \eqref{EQ:calR_EM} as
\begin{align}\label{EQ:calR_EM_as_sum}
    \mathcal{R}_{\mathrm{EM}}(\alpha,\tau)
    &= \sum_{\boldsymbol{\ell}\neq\boldsymbol{0}} \xhat{\mathpzc{z}}^-_{\boldsymbol{\ell}}(\alpha,\tau) - \int_{\mathbb{R}^N} \mathrm{d}\boldsymbol{\ell} \; \xhat{\mathpzc{z}}^-_{\boldsymbol{\ell}}(\alpha,\tau) \cr
    &= \frac{\W e^{\X+\W}}{N!\,G^2(N+1)} \sum_{k=1}^{k_{\text{max}}} Q^{\text{reg}}_{N,k} \, \frac{(2N\Y)^{k}}{k} \; L_{k-1}^1(-\W) \;,
\end{align}
where the coefficients are given by the generalized Euler--Maclaurin expansion\footnote{
    See \cite{Buchheit:2022} for an introduction to generalized Euler--Maclaurin expansions and their physical applications.
}
\begin{align}\label{EQ:Q_reg}
    Q^{\text{reg}}_{N,k} = \sum_{\boldsymbol{\ell}\neq\boldsymbol{0}} \frac{\Delta^2(\ell_1,\ldots,\ell_N)}{|\boldsymbol{\ell}|^{2k}} - \int_{\mathbb{R}^N} \mathrm{d}\boldsymbol{\ell} \; \frac{\Delta^2(\ell_1,\ldots,\ell_N)}{|\boldsymbol{\ell}|^{2k}} \;.
\end{align}

We are left with the residual term \eqref{EQ:calR_0} that we rewrite through \eqref{EQ:z_m_integral_formula} as the residue of an essential singularity,
\begin{align}
    \mathcal{R}_0(\alpha,\tau)
    &= \frac{e^{\X+\W}}{N!\,G^2(N+1)} \; P_N \, \Res_{u=2N\Y}\!\Big(e^{\frac{2N\Y\W}{u-2N\Y}} \; u^{\nu-1} \Big) \cr
    &= \frac{W \, e^{\X+\W}}{N!\,G^2(N+1)} \; P_N \; \frac{(2N\Y)^{\nu}}{\nu} \; L_{\nu-1}^1(-\W) \;,
\end{align}
in terms of the constant of group-theoretic origin
\begin{align}
    P_N
    &= \pi^2 \, \int_{\mathbb{R}^N} \mathrm{d}\mflux \; \V \bigg[ |\mflux|^{1-\frac{N}{2}} \, \mathrm{i}^{\frac{N}{2}} \, H^{(1)}_{\frac{N}{2}-1}(2\mathrm{i}\pi|\mflux|) \bigg] \cr
    &= \frac{G(N+2)\,\Gamma(N/2)}{(-2)^{\nu}\,\Gamma(N^2/2)} \;.
\end{align}

By taking advantage of certain cancellations, we can write the deformed partition function in a more suggestive way as
\begin{align}\label{EQ:Z_tau_negative}
    Z(\alpha,\tau)
    = {}&\mathcal{R}_0(\alpha,\tau) + \sum_{\boldsymbol{\ell}\in\mathbb{Z}^N} \frac{\Delta^2(\ell_1,\ldots,\ell_N)}{N!\,G^2(N+1)} \; \bigg(e^{\X-\frac{\W|\boldsymbol{\ell}|^2}{2N\Y-|\boldsymbol{\ell}|^2}} - e^{\X+\W\vphantom{\frac{|^2}{|^2}}}\bigg) \cr
    {}&- e^{\X+\W}\int_{\mathbb{R}^N} \kern -0.15em \mathrm{d}\boldsymbol{\ell} \; \frac{\Delta^2(\ell_1,\ldots,\ell_N)}{N!\,G^2(N+1)} \; \W \sum_{k=1}^{k_{\text{max}}}\frac{1}{k}\left(\frac{2N\Y}{|\boldsymbol{\ell}|^2}\right)^{\!k}\,L_{k-1}^1(-\W) \;.
\end{align}
Here, the counterterms have the same form as the ones that appear in \eqref{EQ:Z_neg_final}, but are provided by an integral over $\boldsymbol{\ell}$, rather than a sum. In fact, in this expression both the sum and the integral are separately divergent. Once more, convergence should be understood as in Footnote~\ref{FN:cesaro_limit}.

The sum over $\boldsymbol{\ell}$ is the only term that survives in the abelian theory, and correctly reproduces the result obtained in \cite{Griguolo:2022xcj}, where a constant is subtracted from the $e^{-aH}$ term associated with the deformed Hamiltonian \eqref{EQ:deformed_Hamiltonian}. The same constant factor, namely
\begin{align}
    e^{\X+\W} = e^{\frac{N^2\alpha}{2\tau}} \;,
\end{align}
appears in front of the integral and in $\mathcal{R}_0$, and determines the nonperturbative character of every counterterm in \eqref{EQ:Z_tau_negative}.

So far, we have been able to recast the full partition function from a sum over fluxes into an expression that is the direct deformation of the sum over $\mathrm{U}(N)$ representations in \eqref{EQ:Z_as_sum_over_els}. We now wish to provide a more compact way to reorganize the result, which should shed more light on the origin of the conterterms.
To this end, we observe that in \eqref{EQ:zhatplus}, instead of subtracting from the exponential the first $k_{\text{max}}+1$ terms of its $1/|\boldsymbol{\ell}|$ expansion, we can equivalently use the rest of the series to write
\begin{align}
    \sum_{\boldsymbol{\ell}\neq\boldsymbol{0}} \xhat{\mathpzc{z}}^{\!+}_{\boldsymbol{\ell}}(\alpha,\tau)
    &= \frac{\W e^{\X+\W}}{N!\,G^2(N+1)} \sum_{k=k_{\text{max}}+1}^{\infty} Q^{\vphantom{\text{g}}}_{N,k} \, \frac{(2N\Y)^k}{k} \; L_{k-1}^1(-\W) \;,
\end{align}
in a way that is similar to the residual term in \eqref{EQ:calR_EM_as_sum}, if not for the fact that the coefficients
\begin{align}
    Q^{\vphantom{\text{g}}}_{N,k} = \sum_{\boldsymbol{\ell}\neq\boldsymbol{0}} \frac{\Delta^2(\ell_1,\ldots,\ell_N)}{|\boldsymbol{\ell}|^{2k}}
\end{align}
are not quite the same as in \eqref{EQ:Q_reg}. However, as explained in Appendix~\ref{APP:hadamard}, $Q^{\text{reg}}_{N,k}$ is the meromorphic continuation of $Q^{\vphantom{\text{g}}}_{N,k}$ in $k$, and both can be expressed as certain derivatives of the regularized Epstein zeta function.
Therefore, we can use the prescription in \eqref{EQ:sum_from_epstein_reg} to write the full partition function as
\begin{align}
    Z(\alpha,\tau) = \mathcal{R}_0(\alpha,\tau) + \frac{\W e^{\X+\W}}{N!\,G^2(N+1)} \sum_{k=1}^{\infty} \frac{(2N\Y)^k}{k} \; L_{k-1}^1(-\W) \; \mathbf{V} \, \EpsZreg{\mathbf{0}}{\mflux}(2k) \, \Big|_{\mflux=0} \;. 
\end{align}
The range of the sum can be safely extended to $k=0$ since the added term vanishes identically.
According to this form of the partition function, the counterterms can be seen as originating from the regularization of the Dirichlet-like sums generated by the expansion in inverse powers of $|\boldsymbol{\ell}|$. There is only a finite number of such terms for any $N$, namely those with $1 \leq k \leq k_{\text{max}}$. The expansion necessarily misses the term with $\boldsymbol{\ell}=\mathbf{0}$, which is accounted for by the presence of $\mathcal{R}_0$. This term vanishes in the undeformed theory due to the presence of the Vandermonde determinant, but amounts to a finite nonperturbative contribution in the deformed theory as prescribed by the analysis of the deformed flux sectors.

\section{The semiclassical limit}\label{SEC:semiclassical_limit}
In Section~\ref{SEC:any_zm_negative}, we commented on the fact that in writing \eqref{EQ:Z_n_negative_tau_odd_N}, the partition function of a generic deformed flux sector for $\tau<0$, we were confronted with a choice regarding the nonperturbative part of the expression. 
At the end of Appendix~\ref{APP:identity}, we noticed that our choice guarantees the convergence of the sum over the fluxes $\mflux$. Although this property is certainly necessary for the validity of our construction, it is not sufficient to fully remove the ambiguity regarding the exact form of the deformed flux sectors.
From the point of view of the flow equation \eqref{EQ:Z_flow_equation}, this arbitrariness comes down to the fact that imposing a boundary condition at $\tau = 0$ is not enough to guarantee the uniqueness of the solution.
A second boundary condition can be imposed by taking $\alpha\to0^+$. However, one should be careful in taking such a limit, since there are a priori various ways in which this can be done. Specifically, keeping $\tau$ finite, or alternatively, keeping $\mu$ finite, both lead to unphysical regimes.

In pure undeformed Yang--Mills theory, taking the $g_{\text{YM}}\to0$ limit is equivalent to taking the semiclassical limit. The reason for this is that the gauge coupling acts as an overall constant that multiplies the action \eqref{EQ:S_YM_with_F}, playing a role analogous to that of $\hbar$. As a consequence, when $g_{\text{YM}}\to0$, the path integral localizes on the field configuration that minimize the Euclidean action.
However, this feature is not shared by the deformed action, as it can be seen from \eqref{EQ:flow_equation_lagrangian} where the two sides of the equation carry different powers of $\mathscr{L}$, and thus of $g_{\text{YM}}$. One should rather define a rescaled deformation parameter $\sigma \sim \mu/g_{\text{YM}}^2$ so that the deformed Lagrangian density \eqref{EQ:deformed_lagrangian} depends on the gauge coupling through an overall power.

In terms of the variables at hand, then, the semiclassical limit amounts to taking $\alpha\to0$ and $\tau\to0$ simultaneously in such a way that
\begin{align}
    \sigma = \frac{4\pi^2}{N} \frac{\tau}{\alpha^2}
\end{align}
is kept fixed. Concretely, throughout this section we will replace $\tau$ with its expression in terms of $\alpha$ and $\sigma$, so that the semiclassical limit of $\mathpzc{z}_{\mflux}$ is simply obtained by studying the regime where $\alpha\to0$.

By performing the limit at the level of each individual flux sector we expect to find
\begin{align}
    - \log \mathpzc{z}_{\mflux} \sim S_{\text{cl}}(\mflux,\sigma) \;,
\end{align}
where
\begin{align}\label{EQ:S_cl_flux_deformed}
    S_{\text{cl}}(\mflux,\sigma)
    &= \frac{3\pi^2N}{2\alpha\sigma}\left({}_3F_2\mleft(-\frac{1}{2},-\frac{1}{4},\frac{1}{4};\frac{1}{3},\frac{2}{3};\frac{256}{27}|\mflux|^2\sigma\mright)-1\right)
\end{align}
is the deformed action evaluated on the classical instanton configuration, obtained by plugging \eqref{EQ:S_cl_undeformed} into \eqref{EQ:deformed_lagrangian}. Notice that, since \eqref{EQ:deformed_lagrangian} is a strictly monotonic function of the undeformed Lagrangian density, the classical instanton configurations are stationary points of the deformed action as well. 

The remainder of this section is devoted to the computation of the semiclassical limit of $\mathpzc{z}_{\mflux}$ for both sign choices of $\tau$.

\subsection{\texorpdfstring{$\tau>0$}{tau > 0}}
Choosing a positive deformation parameter corresponds to imposing $\sigma>0$. We start with \eqref{EQ:zm_pos_integral_formula} and notice that the integrand is an even function of $r$. Thus we can rewrite the expression by mirroring the integration range about $r=0$ and dividing by 2. We then change integration variable with
\begin{align}
    r = \frac{2\pi N|\mflux|}{\alpha} \, w \;.
\end{align}
The integral now extends over the range $(-w_{\text{sing}},w_{\text{sing}})$, with \begin{align}
    w_{\text{sing}} = \frac{1}{|\mflux|\sqrt{\sigma}} \;.
\end{align}
At the endpoints of such a range, the integrand has an essential singularity.
 
\begin{figure}[tbp]
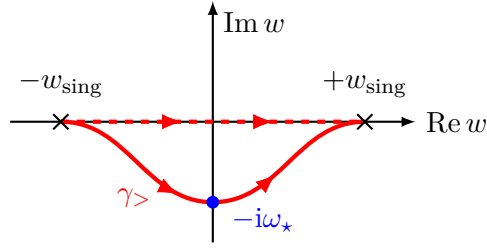

    \centering\saddleposfigure
    \caption{\label{FIG:saddle_pos_contour} The integration contour for \eqref{EQ:zm_pos_integral_formula_classical_limit}. The original integration contour is represented with a dashed red line and runs along the real $w$-axis. In the $\alpha\to0$ limit, we deform the contour according to the steepest-descent approximation which prescribes crossing the saddle point at $-\mathrm{i}\omega_\star$ horizontally. The new contour $\gamma_>$ is represented with a solid red line.}
\end{figure}

We now consider the $\alpha\to0$ regime. To this end, we use the asymptotic behavior of the Bessel function, namely
\begin{align}\label{EQ:Bessel_asymptotics}
    J_\nu(2z) \sim \frac{\cos(2z-\nu\pi/2-\pi/4)}{\sqrt{\pi z}} \qquad \text{for $z\to\infty$} \;,
\end{align}
and find
\begin{align}\label{EQ:zm_pos_integral_formula_classical_limit}
    \mathpzc{z}_{\mflux}
    &\sim \frac{(2\pi N/\alpha)^{\frac{N+1}{2}}(-1)^{m}}{N!\,G^2(N+1)}\int_{\gamma_>} \! \mathrm{d}w \; (\mathrm{i}w)^{\frac{N-1}{2}} \; \V \bigg[ |\mflux| \, e^{-\frac{4\pi^2N|\mflux|^2}{\alpha}\,\eta(w)} \bigg] \;,
\end{align}
where
\begin{align}
    \eta(w) = \frac{w^2}{2(1 - |\mflux|^2 \sigma w^2)} + \mathrm{i}w \;.
\end{align}
The trigonometric function in \eqref{EQ:Bessel_asymptotics} was turned into a complex exponential in \eqref{EQ:zm_pos_integral_formula_classical_limit} by adding an appropriate odd function of $w$ that gets cancelled under integration. The integration contour $\gamma_>$ is depicted in Figure~\ref{FIG:saddle_pos_contour}.

We employ the steepest-descent approximation method. The solutions of the saddle point equation, that we write as $\eta'(-\mathrm{i}\omega) = 0$, are captured by the quartic
\begin{align}\label{EQ:saddle_point_equation}
    |\mflux|^4\sigma^2\omega^4 + 2|\mflux|^2\sigma\omega^2-\omega+1 = 0 \;.
\end{align}
There is a saddle,
\begin{align}\label{EQ:saddle_omega_star}
    \omega_\star = {}_3F_2\mleft(\frac{1}{2},\frac{3}{4},\frac{5}{4};\frac{4}{3},\frac{5}{3};\frac{256}{27}|\mflux|^2\sigma\mright) \;,
\end{align}
which is smoothly connected to the one of the undeformed theory. Due to the presence of a branch cut in the hypergeometric function, $\omega_\star$ is real for $|\mflux|^2\sigma<27/256$, and so is
\begin{align}
    \frac{1}{\eta''(-\mathrm{i}\omega_\star)} = {}_3F_2\mleft(\frac{3}{4},\frac{5}{4},\frac{3}{2};\frac{4}{3},\frac{5}{3};\frac{256}{27}|\mflux|^2\sigma\mright) > 0 \;.
\end{align}
For this range of parameters, we deform the contour as indicated in Figure~\ref{FIG:saddle_pos_contour} and find
\begin{align}\label{EQ:zm_pos_semiclassical_limit}
    \mathpzc{z}_{\mflux}
    &\sim \frac{(2\pi N/\alpha)^{\frac{N}{2}}(-1)^{m}}{N!\,G^2(N+1)} \; \V \Big[h(|\mflux|^2\sigma) \, e^{-S_{\text{cl}}(\mflux,\sigma)} \Big] \;.
\end{align}
In the above, we have recognized
\begin{align}
    \frac{4\pi^2N|\mflux|^2}{\alpha}\,\eta(-\mathrm{i}\omega_\star) = S_{\text{cl}}(\mflux,\sigma)
\end{align}
to be the deformed classical action \eqref{EQ:S_cl_flux_deformed}, and we have denoted
\begin{align}
    \sqrt{\frac{\omega_\star^{N-1}}{\eta''(-\mathrm{i}\omega_\star)}} = h(|\mflux|^2\sigma) \;,
\end{align}
where
\begin{align}\label{EQ:fluctuations_classical_limit}
    h(z)
    &= \left[{}_3F_2\mleft(\frac{1}{2},\frac{3}{4},\frac{5}{4};\frac{4}{3},\frac{5}{3};\frac{256}{27}z\mright)\right]^{\frac{N-1}{2}} \left[{}_3F_2\mleft(\frac{3}{4},\frac{5}{4},\frac{3}{2};\frac{4}{3},\frac{5}{3};\frac{256}{27}z\mright)\right]^{\frac{1}{2}} \cr 
    &= 1 + (N+2)z + O(z^2) \;.
\end{align}

We can observe that the full result for the deformed flux sector in the semiclassical limit maintains the form in \eqref{z_m_as_w_times_exp}, i.e.\ it can be decomposed as
\begin{align}
    \mathpzc{z}_{\mflux} \sim w_{\mflux}(\alpha,\sigma) \; e^{-S_{\mathrm{cl}}(\mflux,\sigma)} \;.
\end{align}
This is because the action of the differential operator $\V$ in \eqref{EQ:zm_pos_semiclassical_limit} does not spoil the presence of an overall exponential term associated with the deformed classical action. Rather, by acting as in \eqref{EQ:zm_pos_semiclassical_limit}, $\V$ determines the deformation of the fluctuation term $w_{\mflux}(\alpha,\sigma)$.

Notice that for each flux sector there exists a finite neighborhood of $\sigma = 0$ for which $\mathpzc{z}_{\mflux}$ is analytic in $\sigma$. Conversely, for any $\sigma>0$ only a finite number flux sectors will have an associated partition function that is analytic at that point. This reflects the fact that, as discussed in Section~\ref{SEC:Z_tau_positive}, the total partition function has peculiar analyticity properties in any neighborhood of $\tau=0$. The presence of a branch cut can also be understood by noticing that at $|\mflux|^2\sigma=27/256$ the saddle \eqref{EQ:saddle_omega_star} collides with another real solution of \eqref{EQ:saddle_point_equation}. This additional saddle, which is subdominant in the range $0<|\mflux|^2\sigma<27/256$, should combine with the contribution coming from $\omega_\star$ and modify the asymptotic behavior in \eqref{EQ:zm_pos_semiclassical_limit} when $|\mflux|^2\sigma>27/256$.

From the point of view of the full partition function, this feature suggests a semiclassical mechanism for the truncation of the spectrum that can be attributed to a collective behavior of the flux sectors: for any fixed $\tau$, the sum over the fluxes should include contributions coming from both saddles when $|\mflux|$ is large enough. Accordingly, an infinite number of oscillatory terms appears in the full sum, since the saddle-points are complex-conjugates and carry, in general, a non-trivial imaginary part. We argue that a destructive interference occurs among these terms, resulting into the sharp cut-off on the sum over the representations. At a first look, this observation might seem at tension with the fact that the truncation of the spectrum is controlled by the sole $\tau$, given that the location of the branch-point of the classical action depends on both $\tau$ and $\alpha$. We expect nevertheless that the interference should come from the full tower of fluxes, and as such to be dominated by terms with large $|\mflux|$. Nicely, one finds that in said regime, $|S_{\text{cl}}|^2 \sim 4\pi^2|\mflux|^2N^3/\tau$, recovering the dependence on the correct cutoff scale, including the expected power of $N$.

\subsection{\texorpdfstring{$\tau<0$}{tau < 0}}
For negative values of the deformation parameter, i.e.\ for $\sigma<0$, we start from \eqref{EQ:z_m_integral_formula} and rescale the integration variable with
\begin{align}
    u = \frac{4\pi^2N^2|\mflux|^2}{\alpha^2} \, v \;.
\end{align}
The essential singularity of the integrand sits now at
\begin{align}
    v_{\text{sing}} &= \frac{1}{|\mflux|^2\sigma} \;.
\end{align}

\begin{figure}[tbp]
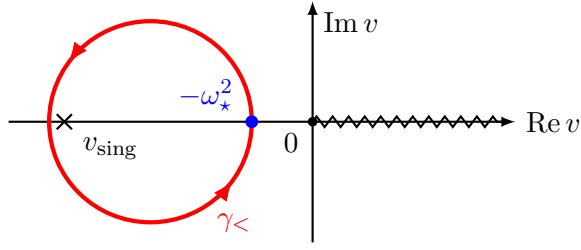

    \centering\saddlenegfigure
    \caption{\label{FIG:saddle_neg_contour} The integration contour for the steepest-descent approximation of \eqref{EQ:classical_limit_saddle_integral} in the $\alpha\to0$ limit. The contour crosses the saddle point $-\omega_\star^2$ parallel to the imaginary axis. Notice that, when approaching $v_{\text{sing}}$ from the left, $\operatorname{Re}\chi\to+\infty$.}
\end{figure}

In taking the $\alpha\to0$ limit, we make use of \eqref{EQ:Hankel_function_asymptotics} to write
\begin{align}\label{EQ:classical_limit_saddle_integral}
    \mathpzc{z}_{\mflux}(\alpha,\tau)
    &\sim \frac{(2\pi N/\alpha)^{\frac{N+1}{2}}}{N!\,G^2(N+1)} \frac{\mathrm{i}}{2} \oint_{\gamma_<} \frac{\mathrm{d}v}{v} \; (\sqrt{-v})^{\frac{N+1}{2}} \; \V \bigg[ |\mflux| \, e^{-\frac{4\pi^2N|\mflux|^2}{\alpha}\,\chi(v)} \bigg] \,,
\end{align}
where
\begin{align}
    \chi(v) = \frac{v}{2(1 - |\mflux|^2 \sigma v)} + \sqrt{-v} \;.
\end{align}
By writing the saddle-point equation as $\chi'(-\omega^2) = 0$, we find that the solutions are again captured by the quartic in \eqref{EQ:saddle_point_equation} and consider the solution in \eqref{EQ:saddle_omega_star}. Since now $\sigma<0$, the saddle is always real for any range of parameters. Moreover, we find that
\begin{align}
    v_{\text{sing}} & < -\omega_\star^2 < 0 \;.
\end{align}
In choosing the contour for \eqref{EQ:classical_limit_saddle_integral} according to the steepest-descent prescription, we notice that $\chi''(-\omega^2_\star) < 0$.
Accordingly, we define $\gamma_<$ so that it crosses the saddle point vertically as indicated in Figure~\ref{FIG:saddle_neg_contour}.
The final result reads
\begin{align}\label{EQ:zm_neg_semiclassical_limit}
    \mathpzc{z}_{\mflux}
    &\sim \frac{(2\pi N/\alpha)^{\frac{N}{2}}}{N!\,G^2(N+1)} \;
    \V \Big[
        h(|\mflux|^2\sigma) \;
        e^{-S_{\text{cl}}(\mflux,\sigma)}
    \Big] \;,
\end{align}
where, again we find that
\begin{align}
    \frac{4\pi^2N|\mflux|^2}{\alpha}\,\chi(-\omega^2_\star) = S_{\text{cl}}(\mflux,\sigma)
\end{align}
is the deformed classical action \eqref{EQ:S_cl_flux_deformed} evaluated on the classical instanton configuration with total magnetic flux $\mflux$, while
\begin{align}
    \sqrt{-\frac{\omega_\star^{N-3}}{4\chi''(-\omega^2_\star)}} = h(|\mflux|^2\sigma)
\end{align}
coincide with the expression obtained in \eqref{EQ:fluctuations_classical_limit}.

It is remarkable that, for each flux sector, the semiclassical limits obtained for either signs of $\tau$ agree.\footnote{
    The only discrepancy between \eqref{EQ:zm_pos_semiclassical_limit} and \eqref{EQ:zm_neg_semiclassical_limit}, namely the presence of the overall sign $(-1)^m$, is merely due to the fact that for $\tau<0$ this was dropped since it is always trivial for odd $N$.
} This effectively tells us that the nonperturbative corrections included in the partition function \eqref{EQ:Z_n_negative_tau_odd_N} are precisely those that guarantee such a match. In fact, each term that appears in the sum generates an instanton-like contribution of the form
\begin{align}
    e^{\X+\W} = e^{\frac{2\pi^2N}{\alpha\sigma}}
\end{align}
that shapes the semiclassical limit.

\section*{Acknowledgments}
We thank Roberto Tateo for interesting discussions. R.P.\ thanks the Galileo Galilei Institute for hospitality during various stages of the project.

\appendix

\section{Confluent hypergeometric functions}
The confluent hypergeometric differential equation
\begin{align}
    z\,\frac{\mathrm{d}^2w}{\mathrm{d}z^2} + (b-z)\,\frac{\mathrm{d}w}{\mathrm{d}z} - a w = 0 
\end{align}
has solution
\begin{align}
    w = c_1 \; {}_1F_1(a;b;z) + c_2 \; U(a,b,z) \;,
\end{align}
where the functions
\begin{align}
    {}_1F_1(a;b;z) &= \sum_{k=0}^\infty \frac{(a)_k}{(b)_k} \frac{z^k}{k!} \;, \\
    U(a,b,z) &= \frac{\Gamma(1-b)}{\Gamma(a-b+1)}\,{}_1F_1(a;b;z) + \frac{\Gamma(b-1)}{\Gamma(a)}\,z^{1-b}\,{}_1F_1(a-b+1;2-b;z) \;, \label{EQ:U_from_1F1}
\end{align}
are referred to, respectively, as Kummer and Tricomi confluent hypergeometric functions. Notice that ${}_1F_1(a;b;z)$ does not exist when $b$ is a nonpositive integer, and that \eqref{EQ:U_from_1F1} holds for $b$ noninteger. One can extend the definition of the ${}_1F_1$ by using
\begin{align}
    \lim_{b\to-n} \frac{{}_1F_1(a;b;z)}{\Gamma(b)} = \frac{(a)_{n+1}}{(n+1)!} \, z^{n+1} \; {}_1F_1(a+n+1;n+2;z) \;,
\end{align}
for $n$ nonnegative integer.
The function $U(a,b,z)$ has a branch cut in the complex $z$-plane along $z\in(-\infty,0]$. For $x<0$,
\begin{align}
    U(a,b,x+\mathrm{i}0) &= U(a,b,x) \;, \cr
    U(a,b,x-\mathrm{i}0) &= e^{2\mathrm{i}b\pi} \, U(a,b,x) - \frac{2\mathrm{i}\pi e^{\mathrm{i}b\pi}}{\Gamma(a-b+1)\,\Gamma(b)} \, {}_1F_1(a;b;x) \;.
\end{align}

The Kummer confluent hypergeometric function can be recast in terms of generalized Laguerre polynomials as
\begin{align}
    {}_1F_1(a;b;z) &= \frac{\Gamma(1-a)\,\Gamma(b)}{\Gamma(b-a)} \, L_{-a}^{b-1}(z) \;.
\end{align}

Two asymptotic behaviors are particularly relevant for the present work: for $x\to+\infty$,
\begin{align}
    U(a,b,x) &\sim x^{-a} \;, \label{EQ:Tricomi_asymptotics} \\
    {}_1F_1(a;b;-x) &\sim \frac{\Gamma(b)}{\Gamma(b-a)} \, x^{-a} \left(1 + O\mleft(\frac{1}{x}\mright)\right) + e^{-x} \ldots \;. \label{EQ:Kummer_asymptotics}
\end{align}

The confluent hypergeometric functions enjoy the two following integral representations.
For $\operatorname{Re}a>0$ and $\operatorname{Re}z>0$,
\begin{align}\label{EQ:Tricomi_integral_representation}
    U(a,b,z) = \frac{1}{\Gamma(a)} \int_0^{\infty} \mathrm{d}t \; e^{-zt} \, t^{a-1} \, (t+1)^{b-a-1} \;.
\end{align}
For $\operatorname{Re}a>0$,
\begin{align}\label{EQ:Kummer_integral_representation}
    {}_1F_1(a;b;z) = \frac{1}{2\pi\mathrm{i}} \frac{\Gamma(b)\,\Gamma(a-b+1)}{\Gamma(a)} \int_0^{(1+)} \mathrm{d}t \; e^{zt} \, t^{a-1} \, (t-1)^{b-a-1} \;.
\end{align}
The last integral is taken over a contour starting and ending in 0 and encircling 1 in the positive sense.

For $\operatorname{Re}y>1/2$, the two multiplication theorems hold,
\begin{align}
    \sum_{j=0}^\infty \frac{1}{j!}\left(\frac{1}{y}-1\right)^j U(a-j,b,x) &= e^{x(1-y)} y^{b-a} \, U(a,b,xy) \;, \label{EQ:Tricomi_multiplication_theorem} \\
    \sum_{j=0}^\infty \frac{(b-a)_j}{j!}\left(1-\frac{1}{y}\right)^j {}_1F_1(a-j;b;x) &= e^{x(1-y)} y^{b-a} \, {}_1F_1(a;b;xy) \;. \label{EQ:Kummer_multiplication_theorem}
\end{align}

\section{Some useful identities}\label{APP:identity}
\begin{figure}[tbp]
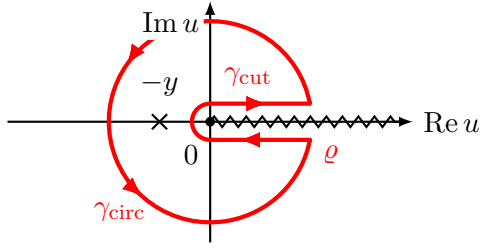

    \centering\contourfigure
    \caption{\label{FIG:contour} The integration contour for \eqref{EQ:Kummer_integral_representation_special}. The integrand has a branch cut along the positive real $u$-axis if $a$ is not integer, and an essential singularity at $u=-y$. The contour $\gamma$ is the sum of a circle $\gamma_{\text{circ}}$ of radius $\varrho>y$ and of a Hankel-like contour $\gamma_{\text{cut}}$ wrapping the cut.}
\end{figure}

In this appendix, we prove various identities that are used in Section~\ref{SEC:Z_tau_negative}. We start from \eqref{EQ:Z_n_negative_tau_odd_N} and \eqref{EQ:p_s_generic}. The sum over $s$ can be performed by using the integral representation
\begin{align}\label{EQ:Kummer_integral_representation_special}
    z\,{}_1F_1(a+1;2;z) = \frac{\mathrm{1}}{2\pi\mathrm{i}} \oint_\gamma \frac{\mathrm{d}u}{u} \; e^{\frac{zu}{u+y}} \, \bigg({-}\frac{y}{u}\bigg)^{\!-a} \;,
\end{align}
which is obtained directly from \eqref{EQ:Kummer_integral_representation} through integration by parts and with the change of integration variable
\begin{align}
    t &= \frac{u}{u+y} \;,
\end{align}
with $y>0$.
The integral is performed over the contour $\gamma$ depicted in Figure~\ref{FIG:contour}.

For our purposes, it is convenient to use $y = -2NY$. This gives\footnote{
    The absence of the $(-1)^m$ factor coming from \eqref{EQ:p_s_generic} is merely due to the fact that $m$ is always even if $N$ is chosen to be odd.
}
\begin{align}
    \mathpzc{z}^+_{\mflux}(\alpha,\tau)
    &= \frac{(-1)^{\nu}\pi^{\frac{N}{2}}e^{\X}}{N!\,G^2(N+1)} \frac{\mathrm{i}}{2} \oint_\gamma \frac{\mathrm{d}u}{u} \; e^{\frac{u\W}{u-2N\Y}} \, (-u)^{\frac{N^2}{2}} \; \V \bigg[ (\pi^2|\mflux|^2)^{\nu}\cr
    &\quad \times \bigg({}_1\tilde{F}_2\Big(1;\tfrac{N^2}{2},\nu+1;-\pi^2|\mflux|^2u\Big) - \pi|\mflux|\sqrt{-u} \; {}_1\tilde{F}_2\Big(1;\tfrac{N^2+1}{2},\nu+\tfrac{3}{2};-\pi^2|\mflux|^2u\Big)\bigg)\bigg] \cr \label{EQ:zplus_integral_reprentation} \\
    \mathpzc{z}^-_{\mflux}(\alpha,\tau)
    &= \frac{(-1)^{\nu}\pi^{\frac{N}{2}}e^{\X}}{N!\,G^2(N+1)} \frac{\mathrm{i}}{2} \oint_\gamma \frac{\mathrm{d}u}{u} \; e^{\frac{u\W}{u-2N\Y}} \, (-u)^{\frac{N^2}{2}} \cr
    &\quad \times \V \bigg[ {-}(\pi^2|\mflux|^2)^{\nu} \sum_{s = 1-\frac{N^2}{2}}^{-1/2} \frac{(-\pi^2|\mflux|^2u)^{s}}{\Gamma(s+1+\nu)\,\Gamma(s+N^2/2)} \bigg] \;. \label{EQ:zminus_integral_reprentation}
\end{align}
Notice that the terms obtained by summing over half-integer $s$'s in \eqref{EQ:Z_n_negative_tau_odd_N}, namely \eqref{EQ:zminus_integral_reprentation} and the term with the second regularized hypergeometric in \eqref{EQ:zplus_integral_reprentation}, are free from branch cuts along the positive real $u$-axis. These are, in fact, the terms associated with an integer $a$ in \eqref{EQ:Kummer_integral_representation_special}.

Let us focus first on $\mathpzc{z}^+_{\mflux}(\alpha,\tau)$, with the goal of proving Eq.~\eqref{EQ:zplus_identity}. We separate the contributions coming from $\gamma_{\text{cut}}$ and $\gamma_{\text{circle}}$ and denote them as $\mathpzc{z}^+_{\mflux}(\alpha,\tau) = I_{\text{cut}} + I_{\text{circ}}$.
As mentioned, the integral over the Hankel-like contour receives contributions only from the first regularized hypergeometric. We use \eqref{EQ:1F2_identity_1} and the same argument of Section~\eqref{SEC:Z_tau_positive} to write
\begin{align}\label{EQ:I_cut_1}
    I_{\text{cut}}
    &= \frac{\pi e^{\X}}{N!\,G^2(N+1)} \frac{\mathrm{i}}{2} \int_{\gamma_{\text{cut}}} \! \frac{\mathrm{d}u}{u} \; e^{\frac{u\W}{u-2N\Y}} \sqrt{-u}
    \; \V \bigg[|\mflux|^{1-\frac{N}{2}} \, u^{\frac{N}{4}} \, J_{\frac{N}{2}-1}(2\pi|\mflux|\sqrt{u})\bigg] \;.
\end{align}
The counterterms that appear in \eqref{EQ:zplus_identity} come from the integral over the circle. We can split the exponential in the integrand in terms of its power expansion in $1/u$ as
\begin{align}
    e^{\frac{u\W}{u-2N\Y}} = \mathcal{E}_1 + \mathcal{E}_2 \;,
\end{align}
where
\begin{align}
    \mathcal{E}_1
    &= e^\W + \W e^\W \sum_{k=1}^{k_{\text{max}}}\frac{1}{k}\left(\frac{2N\Y}{u}\right)^{\!k}\,L_{k-1}^1(-\W) \;, \\
    \mathcal{E}_2
    &= \W e^\W \sum_{k=k_{\text{max}}+1}^\infty\frac{1}{k}\left(\frac{2N\Y}{u}\right)^{\!k}\,L_{k-1}^1(-\W) \;,
\end{align}
and $k_{\text{max}} = (N^2-1)/2$. With an obvious notation we denote $I_{\text{circ}} = I_{\text{circ},1} + I_{\text{circ},2}$.
We notice that in
\begin{align}
    I_{\text{circ},1}
    &= \frac{\pi e^{\X}}{N!\,G^2(N+1)} \frac{\mathrm{i}}{2} \int_{\gamma_{\text{circ}}} \! \frac{\mathrm{d}u}{u} \; \mathcal{E}_1 \, \sqrt{-u}
    \; \V \bigg[|\mflux|^{1-\frac{N}{2}} \, u^{\frac{N}{4}} \, J_{\frac{N}{2}-1}(2\pi|\mflux|\sqrt{u})\bigg] \cr
    &\quad + \frac{(-1)^{\nu}\pi^{2\nu+\frac{N}{2}+1}e^{\X}}{N!\,G^2(N+1)} \frac{\mathrm{i}}{2} \int_{\gamma_{\text{circ}}} \!\! \mathrm{d}u \; \mathcal{E}_1 \, u^{k_{\text{max}}} \, \V \bigg[|\mflux|^{2\nu+1} \, {}_1\tilde{F}_2\Big(1;\tfrac{N^2+1}{2},\nu+\tfrac{3}{2};-\pi^2|\mflux|^2u\Big)\bigg] \cr
\end{align}
the second line vanishes, as it is the integral of an analytic function over a closed contour. Moreover, we can use the fact that now the integrand in the first line does not have an essential singularity at $u = 2N\Y$ to shrink the contour around the cut. In doing so, one can combine the above with \eqref{EQ:I_cut_1} to generate the appropriate counterterms. In taking $\varrho\to\infty$, we see that $I_{\text{cut}}+I_{\text{circ},1}$ reproduces \eqref{EQ:zplus_identity}. What is left to show now is that, in the same limit, the remaining term $I_{\text{circ},2}$ vanishes.
To this end, we use
\begin{align}\label{EQ:1F2_identity_2}
    {}_1\tilde{F}_2\Big(1;\tfrac{N^2+1}{2},\nu+\tfrac{3}{2};-z^2\Big) = z^{\frac{N}{2}-N^2} \, J_{1-\frac{N}{2}}(2z) + \sum_{s=1-\frac{N^2}{2}}^{-\frac{1}{2}} \frac{(-z^2)^{s-\frac{1}{2}}}{\Gamma(s+\nu+1)\,\Gamma(s+N^2/2)}  \;.
\end{align}
The two Bessel functions combine in a Hankel function of the first kind with
\begin{align}
    (-1)^\nu J^{\vphantom{(}}_{\frac{N}{2}-1}(2z) + \mathrm{i} J^{\vphantom{(}}_{1-\frac{N}{2}}(2z)
    &= (-1)^{\nu} \, H^{(1)}_{\frac{N}{2}-1}(2z) \;.
\end{align}
The asymptotic behavior of the Hankel function, namely
\begin{align}\label{EQ:Hankel_function_asymptotics}
    H^{(1)}_{\nu}(2z) &\sim \textstyle{\sqrt{1/(\pi z)}} \, e^{\mathrm{i}(2z-\nu\pi/2-\pi/4)} \qquad \text{for $z\to\infty$} \;,
\end{align}
is sufficient to show that
\begin{align}
    I_{\text{circ},2}
    &= \frac{\pi^{N/2}e^{\X}}{N!\,G^2(N+1)} \frac{\mathrm{i}}{2} \int_{\gamma_{\text{circ}}} \!\! \frac{\mathrm{d}u}{u} \; \mathcal{E}_2 \, \sqrt{-u} \cr
    &\quad \times \V \bigg[ (\pi|\mflux|)^{1-\frac{N}{2}} (\mathrm{i}\sqrt{-u})^{\frac{N}{2}} \, H^{(1)}_{\frac{N}{2}-1}(2\mathrm{i}\pi|\mflux|\sqrt{-u}) + \! \sum_{s=1-\frac{N^2}{2}}^{-\frac{1}{2}} \! \frac{(-\pi^2|\mflux|^2)^{s+\nu} \, u^{s+k_{\text{max}}}}{\Gamma(s+\nu+1)\,\Gamma(s+N^2/2)} \bigg] \cr
\end{align}
vanishes in the $\varrho\to\infty$ limit for the Jordan's lemma. This concludes the proof of \eqref{EQ:zplus_identity}.

We are now in the position to comment on the convergence of the sum over all flux sectors at $\tau<0$. In fact, from \eqref{EQ:1F2_identity_2} we see that $\mathpzc{z}^-_{\mflux}(\alpha,\tau)$ cancels the finite sum generated by the second regularized hypergeometric in $\mathpzc{z}^+_{\mflux}(\alpha,\tau)$ and one is left with an expression with the sole Hankel function, namely
\begin{align}\label{EQ:z_m_integral_formula}
    \mathpzc{z}_{\mflux}(\alpha,\tau)
    &= \frac{\pi \, e^{\X}}{N!\,G^2(N+1)} \frac{\mathrm{i}}{2} \oint_\gamma \frac{\mathrm{d}u}{u} \; e^{\frac{u\W}{u-2N\Y}} \sqrt{-u} \; \V \bigg[ |\mflux|^{1-\frac{N}{2}} \, (\mathrm{i}\sqrt{-u})^{\frac{N}{2}} \, H^{(1)}_{\frac{N}{2}-1}(2\mathrm{i}\pi|\mflux|\sqrt{-u}) \bigg] \,.
\end{align}
By closing the contour $\gamma$ around the essential singularity at $u=2NY$, the Hankel function generates, according to \eqref{EQ:Hankel_function_asymptotics}, an exponential term of the form $e^{-2\pi|\mflux|\sqrt{-2N\Y}}$ that ensures the convergence of the sum over $\mflux$.

\section{Some useful tools}
\label{APP:hadamard}
In this appendix, we follow the notation of \cite{Buchheit:2022}.

Let us consider the $N$-dimensional integral
\begin{align}
    I_{\nu,\varepsilon} = \int\limits_{\Omega\setminus B_{\varepsilon}} \!\! \mathrm{d}\mathbf{x} \; \frac{f(\mathbf{x})}{|\mathbf{x}|^\nu} \;,
\end{align}
where $\Omega$ is some region of $\mathbb{R}^N$ that contains the origin, and $f$ is sufficiently differentiable. The integral $I_{\nu,0}$ converges for $\Re\nu<N$. The \emph{Hadamard finite-part integral}
\begin{align}
    \fint_\Omega \mathrm{d}\mathbf{x} \; \frac{f(\mathbf{x})}{|\mathbf{x}|^\nu} = \lim_{\varepsilon\to 0}\bigg(I_{\nu,\epsilon} - \mathbf{H}_{\nu,\varepsilon}f(\mathbf{0})\bigg) \;,
\end{align}
is the analytic continuation of $I_{\nu,0}$ to $\nu \in \mathbb{C}\setminus (\mathbb{N}+N)$ defined through the subtraction generated the differential operator
\begin{align}
    \mathbf{H}_{\nu,\varepsilon} =
    \sum_{n=0}^{\lfloor\Re\nu-N\rfloor} \frac{1}{n!} \int\limits_{\mathbb{R}^d\setminus B_\varepsilon} \!\! \mathrm{d}\mathbf{x} \; \frac{(\mathbf{x}\cdot\boldsymbol{\nabla})^n}{|\mathbf{x}|^\nu} \;.
\end{align}
For $\nu \in (\mathbb{N}+N)$, we define instead
\begin{align}
    \mathbf{H}_{\nu,\varepsilon} = \sum_{n=0}^{\nu-N-1} \frac{1}{n!} \int\limits_{\mathbb{R}^d\setminus B_\varepsilon} \!\! \mathrm{d}\mathbf{x} \; \frac{(\mathbf{x}\cdot\boldsymbol{\nabla})^n}{|\mathbf{x}|^\nu} + \frac{1}{(\nu-N)!} \int\limits_{B_1\setminus B_\varepsilon} \!\! \mathrm{d}\mathbf{x} \; \frac{(\mathbf{x} \cdot \boldsymbol{\nabla})^{\nu-N}}{|\mathbf{x}|^\nu} \;,
\end{align}
which amounts to dropping logarithmic and power-law divergences in $I_{\nu,\varepsilon}$.

The \emph{Epstein zeta function}
\begin{align}
    \EpsZ{\mathbf{y}}{\mathbf{p}}(\nu) = \sum_{\mathbf{x}\neq\mathbf{y}} \frac{e^{-2\pi\mathrm{i}\mathbf{p}\cdot\mathbf{x}}}{|\mathbf{x}-\mathbf{y}|^{\nu}} \;,
\end{align}
is smooth in $\mathbf{p}\in\mathbb{R}^N\setminus\mathbb{Z}^N$. By subtracting the singularity in $\mathbf{p}=\mathbf{0}$ one can define the regularized Epstein zeta function
\begin{align}
    \EpsZreg{\mathbf{y}}{\mathbf{p}}(\nu) = e^{2\pi\mathrm{i}\mathbf{p}\cdot\mathbf{y}} \; \EpsZ{\mathbf{y}}{\mathbf{p}}(\nu) - \frac{\Gamma(N/2-\nu/2)}{\pi^{N/2-\nu}\,\Gamma(\nu/2)} \, |\mathbf{p}|^{N-\nu} \;,
\end{align}
which is analytic in $\mathbf{p}=\mathbf{0}$.

Let us now consider a polynomial $P$. The sum
\begin{align}
    S_{\nu} = \sum_{\mathbf{x}\neq\mathbf{0}} \frac{P(\mathbf{x})}{|\mathbf{x}|^\nu}
\end{align}
is well-defined for $\Re(\nu)>N+\deg{P}$. A meromorphic continuation of $S_\nu$ outside the region where it converges is given by \cite{Buchheit:2021}
\begin{align}
    S_\nu = \lim_{\beta\to0} \Bigg( \sum_{\mathbf{x}\neq\mathbf{0}} e^{-\beta|\mathbf{x}|^2} \, \frac{P(\mathbf{x})}{|\mathbf{x}|^\nu} - \fint_{\raisebox{-.3em}{$\scriptstyle\mathbb{R}^N$}} \!\! \mathrm{d}\mathbf{x} \; e^{-\beta|\mathbf{x}|^2} \, \frac{P(\mathbf{x})}{|\mathbf{x}|^\nu} \Bigg) \;.
\end{align}
Alternatively, the same continuation can be obtained by taking derivatives of the regularized Epstein zeta function, i.e.\
\begin{align}\label{EQ:sum_from_epstein_reg}
    S_{\nu} = P\mleft(\frac{\mathrm{i}\boldsymbol{\nabla}_\mathbf{\!p}}{2\pi}\mright) \; \EpsZreg{\mathbf{0}}{\mathbf{p}}(\nu) \, \Big|_{\mathbf{p}=0} \;.
\end{align}

\bibliographystyle{JHEP}
\bibliography{bibliography}
\end{document}